\documentclass[12pt]{article}%
\usepackage{amssymb}
\usepackage{amsmath}
\usepackage{amsfonts}
\usepackage{graphicx}%
\begin{document}
\setcounter{page}1 \pagestyle{plain}

\noindent{\small usc-hep/05020b1\hfill\hfill hep-th/0502065}\newline%
{\small \hfill}

{\vskip 0.7cm}

\begin{center}
{\Huge Twistors and 2T-physics}{\small \footnote{{\small Research supported by
the US Department of Energy, Grant No. DE-FG03-84ER40168}}}

{\vskip 0.5cm}

\textbf{Itzhak Bars}

{\vskip 0.4cm}

\textbf{Department of Physics and Astronomy}

\textbf{University of Southern California}

\textbf{Los Angeles, CA 90089-0484}

{\vskip 0.5cm}

\textbf{ABSTRACT}
\end{center}

\noindent Two-Time physics applies broadly to the formulation of physics and
correctly describes the physical world as we know it. Recently it was applied
to a 2T re-formulation of the $d=4$ twistor superstring, which was suggested
by Witten as an efficient approach for computations of physical processes in
the maximally supersymmetric $N=4$ Yang-Mills field theory in four dimensions.
The 2T formalism provides a six dimensional view of this theory and suggests
the existence of other $d=4$ dual forms of the same theory. Furthermore the 2T
approach led to the first formulation of a twistor superstring in $d=10$
appropriate for AdS$_{5}\times$S$^{5}$ backgrounds, and a twistor superstring
in $d=6$ related to the little understood superconformal theory in $d=6.$ The
proper generalization of twistors to higher dimensions is an essential
ingredient which is provided naturally by 2T-physics. These developments are
summarized in this lecture{\small \footnote{{\small {Lectures delivered at}
\textquotedblleft Twistor String Theory", Oxford, England, Jan. 2005; and  }
\par
{\small \quad\textquotedblleft Fundamental Interactions and Twistor Methods",
Wroclaw, Poland, Oct. 2004.  }
\par
{\small \quad{Transparencies available at http://physics.usc.edu/\symbol{126}%
bars/papers/twistor.pdf.}  }
\par
{\small \hspace{0.1in}}}}.

\section{Introduction}

Two-Time Physics (2T-physics) is a natural framework in higher spacetime with
2T signature that encodes and unifies many aspects of physics, from simple
quantum mechanics to strings \cite{2treviews}. The 2T-physics formalism is
free from any problems with unitarity or causality, thanks to appropriate
gauge symmetries. One-time Physics (1T-physics) is correctly embedded in the
2T-physics framework.

The 1T interpretation of a 2T-physics system depends on\ the
\textit{perspective}\ of embedding phase space in $(d-1,1)$ dimensions into
phase space in\ $(d,2)$ dimensions. This is done by making a gauge choice,
which yields a holographic image of the physical subspace of the $(d,2)$ phase
space. Consequently, the same system in $(d,2)$ is viewed as various
holographic dynamical images in various $(d-1,1)$ embeddings. From the point
of view of 2T-physics, many aspects of 1T-physics, such as the Hamiltonian
with interactions, \textit{time and space, are all emergent\ concepts} that
depend on the embedding. In particular \textit{twistors} provide a particular
holographic image of the system in $\left(  d,2\right)  $. The 2T-physics
formalism leads to the proper generalization of twistors to any dimension
\cite{2ttwistor} as will be outlined in this lecture and presented in more
detail in \cite{twistorD}.

The $d=4$ twistor superstring developed by Witten \cite{witten,witten2} and
Berkovits \cite{berko,berko2,berkWit} coincide with a 2T superstring
\cite{2tsuperstring} in $\left(  4,2\right)  $ dimensions with SU$\left(
2,2|4\right)  $ supersymmetry, when the 2T superstring is discussed from the
perspective of twistors. In addition, the 2T superstring approach in $\left(
6,2\right)  $ dimensions with OSp$\left(  8|2\right)  $ supersymmetry, and
$\left(  10,2\right)  $ dimensions with SU$\left(  2,2|4\right)  $
supersymmetry, yielded new twistor superstrings that were conceived for the
first time, thus demonstrating the usefulness of the 2T-physics formalism. The
$\left(  10,2\right)  $ case yields a holographic twistor description of the
space AdS$_{5}\times\,$S$^{5},$ with a twistor superstring whose particle
limit spectrum is the full Kaluza-Klein towers of type IIB supergravity
compactified on AdS$_{5}\times$S$^{5}$. This spectrum contains information
about hidden dimensions with $(10,2)$ signature as discussed earlier
\cite{2tAdSs}\cite{2tZero}, and the new superstring extends the $\left(
10,2\right)  $ view of AdS$_{5}\times$S$^{5}$ to the realm of strings.
Similarly, the $(6,2)$ case yields a new twistor superstring whose particle
limit describes a supermultiplet of a peculiar self-dual superconformal theory
in $(5,1)$ dimensions whose physical space (in a lightcone gauge) consists of
8 bosons and 8 fermions. Another description \cite{2tsuperstring} of this
supermultiplet is the unitary representation of OSp$\left(  8|4\right)  $ in
the oscillator formalism \cite{barsgunaydin} which coincides with the
OSp$\left(  8|4\right)  $ doubleton given in \cite{gunaydinWarner}. This six
dimensional conformal theory is expected to exist as an interacting theory,
but it cannot be described in the form of a field theory \cite{witten6D}. The
twistor superstring may be a possible description of this interacting theory.

Perhaps I should give some of the history that motivated the development of
2T-physics. It is often stated that 32 supersymmetries is the maximum possible
number of supersymmetries, and therefore 11 dimensions, which has a spinor of
32 real components, is the maximum number of dimensions in a supersymmetric
theory of fundamental physics. However, the Weyl spinor in 12 dimensions, with
signature (10,2), also has a real spinor with 32 components. Furthermore, the
maximally extended supersymmetry algebra, called the M-algebra, has a symmetry
of isomorphisms that include SO$\left(  10,2\right)  ,$ which can be
interpreted as acting on a 12-dimensional spacetime with signature $(10,2)$.
This point of view was expressed for the first time in 1995 and related to
dualities in one of my talks \cite{osaka} and later further developed in
\cite{stheory}. The possibility of hidden timelike dimensions in M-theory was
strengthened further by the hidden symmetry structures in the web of dualities
involving D-branes, as in F-theory \cite{ftheory} and S-theory \cite{stheory}.

Of course hints coming from symmetries, although suggestive, are not enough to
infer extra spacetime dimensions. However, a dynamical theory involving the
higher spacetime, which describes the recognizable world, would go a long way
toward understanding the higher spacetime. This was the motivation behind the
development of 2T-physics, which after some attempts \cite{ibkounnas} finally
took the correct physical form starting in 1998 as described in
\cite{2treviews}. The backbone of the 2T structure is an Sp$\left(
2,R\right)  $ gauge symmetry that acts in phase space. By generalizing this
symmetry in several appropriate ways 2T-physics makes contact with the real
world. By now it is abundantly clear that the 2T framework describes correctly
simple everyday physics as well as complicated structures in string theory.

We don't have to wait until we discover the correct formulation of M-theory to
know that 2T-physics is correct, and that it teaches us that there is a sense
in which $\left(  d,2\right)  $ dimensions provide a higher unifying
framework. This view is already born out in simple classical and quantum
mechanical systems, and there are useful nontrivial consequences that follow
from it. We have now come back full circle to using 2T-physics techniques to
try to construct corners of M-theory, such as the twistor superstrings given
in \cite{2tsuperstring} and described briefly in this lecture. 2T-physics
suggests that we should look for a formulation of M-theory in $\left(
11,2\right)  $ dimensions with global symmetry OSp$\left(  1|64\right)  $
\cite{stheory}\cite{liftM}\cite{2ttoyM}.

In this lecture I will first give a brief description of the concepts in
2T-physics, and then describe the twistors and the twistor superstrings in
$d=3,4,5,6,10$ that were constructed by using the formalism. The twistors that
emerge in the new twistor superstrings can be discussed without the full 2T
formalism, as will be done in part of this lecture and in \cite{twistorD}, so
this aspect can be carried away and applied usefully elsewhere without the
need for the full 2T package. However, the full 2T formalism is what provides
the easy proof that the new twistors in the higher dimensions describe
AdS$_{5}\times\,$S$^{5}$ (for $d=10)$, the six dimensional conformal theory
(for $d=6)$ respectively, and of course the Super Yang Mills (SYM) theory (for
$d=4$). The 2T version of the theory is far richer because it relates the
twistors to other dual forms of the same theory, and this aspect may be
crucial ultimately for deeper understanding and for practical progress.

\section{2T-physics}

\subsection{Gauge symmetry is the origin of spacetime signature}

First I suggest the point of view that gauge symmetry is at the origin of the
1T spacetime signature $\left(  -,+,\ldots,+\right)  ,$ and then show that the
same point of view leads to the 2T signature $\left(  -,-,+,\cdots,+\right)
.$

Consider the action of a particle on the worldline which is invariant under
$\tau$ reparametrizations%
\begin{equation}
S=\int_{1}^{2}d\tau\left(  \partial_{\tau}x^{\mu}p_{\mu}-e(\tau)Q\left(
x,p\right)  ~\right)  , \label{Q1T}%
\end{equation}
where $e(\tau)$ is the gauge field that transforms as $\delta_{\varepsilon
}e(\tau)=\partial_{\tau}\varepsilon\left(  \tau\right)  $, while the
infinitesimal transformations of $x\left(  \tau\right)  ,p\left(  \tau\right)
$ are given by the Poisson brackets $\delta_{\varepsilon}x^{\mu}%
=\varepsilon\left(  \tau\right)  \left\{  Q,x^{\mu}\right\}  =-\varepsilon
\left(  \tau\right)  \partial Q/\partial p_{\mu}$ and $\delta p_{\mu
}=\varepsilon\left(  \tau\right)  \left\{  Q,p_{\mu}\right\}  =\varepsilon
\left(  \tau\right)  \partial Q/\partial x^{\mu}.$ The Lagrangian transforms
into a total derivative so that $\delta_{\varepsilon}S=\int_{1}^{2}%
d\tau\partial_{\tau}\left(  \varepsilon Q\right)  =0,$ with boundary
conditions $\varepsilon\left(  \tau_{1}\right)  =\varepsilon\left(  \tau
_{2}\right)  =0$. The well known free massless relativistic particle action
corresponds to $Q\left(  x,p\right)  =\frac{1}{2}\eta^{\mu\nu}p_{\mu}p_{\nu},$
while the general $Q\left(  x,p\right)  $ can describe all possible
interactions of the particle in any background. For example for an
electromagnetic background we have $Q=\frac{1}{2}\eta^{\mu\nu}\left(  p_{\mu
}-qA_{\mu}\left(  x\right)  \right)  \left(  p_{\nu}-qA_{\nu}\left(  x\right)
\right)  .$ Evidently $Q$ is the generator of the local gauge symmetry. The
equation of motion for $e$ requires $Q\left(  x,p\right)  =0.$ The space in
which the gauge symmetry generator vanishes is evidently gauge invariant (a
singlet under gauge transformations). Therefore, this equation is interpreted
to mean that the physical space (either classical or quantum), defined to be
the solution space of $Q\left(  x,p\right)  =0,$ is gauge invariant.

Consider at first the simplest case $Q\left(  x,p\right)  =p^{2}=0.$ We notice
that if the signature in $p^{2}=\eta^{\mu\nu}p_{\mu}p_{\nu}$ is Euclidean the
only solution of $p^{2}=0$ is $p_{\mu}=0,$ so that no non-trivial solution
exists for physical space for Euclidean signature. To describe non-trivial
motion, target space-time must have 1 time%
\[
p\cdot p=-p_{0}^{2}+\vec{p}^{2}=0.
\]
There are nontrivial solutions also with more timelike dimensions, however
$\tau$ reparametrization is insufficient to remove the ghosts of more than 1
timelike dimension. Therefore unitarity of the theory requires that spacetime
cannot have more than 1 time. Thus, $\tau$ reparametrization requires just one
time coordinate no more and no less. Causality corresponds to admitting only
nonwinding maps $\tau$ $\rightarrow$ $x^{\mu}\left(  \tau\right)  .$

>From the simplest case $Q\left(  x,p\right)  =p^{2}$ we have learned that the
signature of the parameter $\varepsilon\left(  \tau\right)  $ in $\tau$
reparametrization is timelike. Thus, a timelike (or lightlike, but not
spacelike) degree of freedom can be removed from $x^{\mu}\left(  \tau\right)
$ and similarly a timelike degree of freedom can be removed from $p^{\mu}$ by
solving the constraint $p^{2}=0.$ For the more general $Q\left(  x,p\right)  $
the signature of $\varepsilon\left(  \tau\right)  $ is the same as before,
therefore the gauge symmetry will remove a timelike degree of freedom, not a
spacelike one, and the constraint $Q\left(  x,p\right)  =0$ can have a
solution provided target spacetime has signature $\left(  -,+,\ldots,+\right)
$ that includes a timelike degree of freedom. Therefore we deduce that the
gauge symmetry requires that there has to be one timelike degree of freedom in
any target spacetime (relativistic, nonrelativistic, curved, etc.).

This reasoning is broadened by starting with a worldline action that is
invariant under Sp$(2,R)$ gauge symmetry introduced in \cite{2treviews}. For
the simplest case Sp$(2,R)$ acts on phase space as a doublet, and an invariant
action is written as follows%
\begin{gather}
\text{Sp(2,R) doublet}{:}\;\left(
\genfrac{}{}{0pt}{}{X^{M}\left(  \tau\right)  }{P^{M}\left(  \tau\right)  }%
\right)  \equiv X_{i}^{M}\;i=1,2\nonumber\\
A_{i}^{~j}\;\text{gauge field:\ }D_{\tau}X_{i}^{M}=\partial_{\tau}X_{i}%
^{M}-A_{i}^{~j}X_{j}^{M},\;\nonumber\\
S=\frac{\eta_{MN}}{2}\int d\tau(\varepsilon^{ij}\partial_{\tau}X_{i}^{M}%
X_{j}^{N}-A^{ij}~X_{i}^{M}X_{j}^{N})\label{flat2T}\\
\text{Sp(2,R)~generators}\;:X\cdot X,X\cdot P,P\cdot P,\rightarrow\;X_{i}\cdot
X_{j}=0\nonumber
\end{gather}
We deduce that physical space must be Sp(2,R) singlet $X_{i}\cdot X_{j}=0$,
and then ask for which signature $\eta^{{MN}}$ can we find a nontrivial
physical space? We quickly learn that there is no non-trivial content for zero
times or one time, and therefore we must admit that target space-time
$must$\ have 2 times%
\[
-X_{0^{\prime}}^{2}-X_{0}^{2}+X_{I}^{2}=0,\;etc.\;X\cdot P=P\cdot P=0
\]
Compared to $\tau$ reparametrization, Sp(2,R) has 2 more gauge symmetries and
2 more constraints. These eliminate 2 more degrees of freedom from both
$X^{M}$ and $P^{M}$. Thus by starting from a space with signature $\left(
d,2\right)  $ we end up with an emergent spacetime with signature $(d-1,1)$ by
making various gauge choices
\[
(d,2)-(1,1)_{\text{{ gauge parameters}}}^{\text{{ signature of extra~}}%
}=(d-1,1)_{\text{{ space-time}}}^{\text{{ emergent~}}}\text{{ \ }}%
\]
We conclude that physical spacetime has $(d-1)$ space and 1 time, just like
before, but these \textit{must be embedded in a higher spacetime} with
signature $(d,2).$

This would not be very deep if there were a single solution to this embedding.
The non-trivial aspect is that there are many ways in which phase space in
$(d-1,1)$ is embedded in phase space in $(d,2),$ and this provides many ways
in which time (or Hamiltonian) is defined in the emergent spacetime. The
embedding provides a holographic image of the events and motion in the
$\left(  d,2\right)  $ space, which can be interpreted very differently from
the perspective of each of the emergent spaces in $(d-1,1)$ since each such
space defines time (and Hamiltonian) differently than one another. Even though
we start from a single well defined 2T-physics system in $\left(  d,2\right)
,$ we end up with many holographic pictures that are interpreted differently,
with different Hamiltonians, in 1T-physics \cite{2tHandAdS}.

Considering also unitarity we find that no more\ than 2 times are possible
since the Sp$\left(  2,R\right)  $ gauge symmetry cannot remove the ghosts
from a spacetime with more timelike dimensions. In this case causality is
satisfied since the situation in the emergent $(d-1,1)$ is no different than
the one time situation.

The simple model above has been generalized to include arbitrary interactions
with all possible background fields \cite{2tbacgrounds}. The generalized
action is $S=\int d\tau(\partial_{\tau}X^{M}P_{M}-\frac{1}{2}A^{ij}%
Q_{ij}\left(  X,P\right)  ~$and the gauge symmetry is still Sp$\left(
2,R\right)  ,$ with $\delta A_{i}^{~j}=\partial_{\tau}\omega_{i}^{~j}+\left[
A,\omega\right]  _{i}^{~j}$ and $Q_{ij}$ the generator for infinitesimal
transformations $\delta X^{M}=\omega^{ij}\left(  \tau\right)  \left\{
Q_{ij},X^{M}\right\}  =\omega^{ij}\left(  \tau\right)  \partial Q_{ij}%
/\partial P_{M},$ and $\delta P^{M}=\omega^{ij}\left(  \tau\right)  \left\{
Q_{ij},P^{M}\right\}  =-\omega^{ij}\left(  \tau\right)  \partial
Q_{ij}/\partial X^{M}$. The generalized $Q_{ij}\left(  X,P\right)  $ depend on
background fields in $\left(  d,2\right)  $ dimensions, such as $A_{M}\left(
X\right)  ,G_{MN}\left(  X\right)  ,$ etc., and those are constrained by the
requirement that $Q_{ij}\left(  X,P\right)  $ must satisfy the Sp$\left(
2,R\right)  $ algebra. This leads to differential equations for the background
fields. All possible solutions are obtained in \cite{2tbacgrounds}, and it is
shown that this covers all possible interactions with backgrounds in
1T-physics, including the Maxwell field $A_{\mu}\left(  x\right)  ,$ the
gravitational field $g_{\mu\nu}\left(  x\right)  ,$ etc. as described in
Eq.(\ref{Q1T}). A similar but less complete analysis has been done also for
spinning particles \cite{2tfield}.

In this way we can argue that possibly all of 1T-physics can be embedded in
2T-physics. It is evident that from the same 2T-physics model, with fixed
backgrounds, one can obtain in principle many 1T-physics systems in the form
of various holographic images, thus showing that 2T-physics unifies the
various dynamics in 1T into a parent theory in 2T that reveals the hidden
relationships and symmetries that are not evident at all in the 1T approach.

\subsection{Some examples of emergent dynamics and spacetimes}

Consider the simplest case of a 2T-physics action for a particle in flat
$\left(  d,2\right)  $ spacetime as given in Eq. (\ref{flat2T}). In this
lecture I will illustrate two solutions of the constraints that are
inequivalent from the point of view of 1T-physics, but which are evidently
equivalent under the Sp$\left(  2,R\right)  $ gauge transformations of the 2T
theory, and therefore dual to one another. The first case is the relativistic
massless particle and the second case is the hydrogen atom. Sometimes it will
be convenient to express the flat $\left(  d,2\right)  $ metric in lightcone
type basis. By using the extra dimensions $X^{0^{\prime}},X^{1^{\prime}}$ we
define $X^{\pm^{\prime}}=(X^{{0}^{\prime}}\pm X^{{1}^{\prime}})/\sqrt{{2}}$ so
that the metric is $ds^{2}=-\left(  dX^{0^{\prime}}\right)  ^{2}+\left(
dX^{1^{\prime}}\right)  ^{2}++dX^{\mu}dX^{\nu}\eta_{\mu\nu}=-2dX^{+^{\prime}%
}dX^{-^{\prime}}+dX^{\mu}dX^{\nu}\eta_{\mu\nu}$ with $\eta_{\mu\nu}$ the
$\left(  d-1,1\right)  $ Minkowski metric.

\subsubsection{Relativistic spacetime gauge}

In the lightcone type basis we choose two gauges by fixing $X^{+^{\prime}%
}(\tau)=1,~~P^{+^{\prime}}(\tau)=0$ for all $\tau,$ and then solve two of the
constraints $X^{2}=X\cdot P=0.$ This solution is given by%

\begin{align}
X^{M}  &  =\left(  \overset{+^{\prime}}{1},\overset{-^{\prime}}{x^{2}{/2}%
},\overset{\mu}{x^{\mu}}\right)  \;\;{X\cdot X=-2X}^{-^{\prime}}{X}%
^{+^{\prime}}{+X}^{\mu}{X}^{\nu}{\eta}_{\mu\nu}{=0}\label{particlegauge}\\
P^{M}  &  =\left(  \overset{+^{\prime}}{0},\overset{-^{\prime}}{x\cdot
p},\overset{\mu}{p^{\mu}}\right)  \;\;{X\cdot P=0\;,\ P\cdot P}=p^{2}\nonumber
\end{align}
There remains one more gauge choice to be made and one more constraint
$p^{2}=0$ to be solved, but we refrain from doing those steps for now. To
interpret our choice of independent variables $x^{\mu},p^{\mu}$ we investigate
the form of the gauge fixed action%
\[
\text{{ gauge fixed \ }}S=\int d\tau\left(  \dot{x}\cdot p-\frac{1}{2}%
A^{22}p^{2}\right)
\]
and note that $x^{\mu},p^{\mu}$ are canonical variables which describe the
massless relativistic particle in $\left(  d-1,1\right)  $ dimensions. To be
sure of this fact, we can also investigate that the original equations of
motion $\dot{X}^{M}=A^{22}P^{M}+A^{12}X^{M}$ and $\dot{P}^{M}=-A^{11}%
X^{M}-A^{12}P^{M}$ are fully consistent with the equations of motion that
follow from the gauge fixed action.

The original gauge invariant action was symmetric under the SO$\left(
d,2\right)  $ global symmetry$.$ The conserved gauge invariant generators of
that symmetry are $L^{MN}=\varepsilon^{ij}X_{i}^{M}X_{j}^{N}=X^{M}P^{N}%
-X^{N}P^{M}.$ Since both the action and the generators are gauge invariant,
the gauge fixed action must have a hidden SO$\left(  d,2\right)  $ symmetry,
with generators given by the gauge fixed form of $L^{MN}.$ Indeed, the gauge
fixed generators become the conformal symmetry of the massless particle%
\begin{gather*}
\text{gauge fixed }L^{\emph{MN}}\ \text{become\ conformal\ SO}(d,2)\\
L^{\mu\nu}=x^{[\mu}p^{\nu]},\;\;\;L^{+^{\prime}-^{\prime}}=x\cdot p,\,\,\\
L^{+^{\prime}\mu}=p^{\mu},\,\,\;L^{-^{\prime}\mu}=\frac{x^{2}}{2}p^{\mu
}-x\cdot p~x^{\mu}\,.
\end{gather*}
When the system is quantized in terms of the relativistic variables $x^{\mu
},p^{\mu}$ the $L^{MN}$ must be carefully quantum ordered. The ordering must
insure Hermitian $L^{MN}$ relative to a relativistic norm for the quantum
states. When this is done \cite{2treviews} one can compute the Casimir
eigenvalue of the SO$\left(  d,2\right)  $ representation that describes the
massless spinless particle. The result is in complete agreement with covariant
quantization of the 2T system, and is given by
\[
\underset{\text{same as SO}\left(  d,2\right)  \text{ }{covariant}\text{
quantization in Sp(2,R) invariant space}}{\text{{After\ quantum\ ordering:}%
}{~}C_{2}=\frac{1}{2}L^{MN}L_{MN}=1-\frac{d^{2}}{4}}%
\]
This representation is known as the singleton in $d=3,4$ and thus we will call
it the singleton for any $d.$ Note that at the classical level (not watching
orders of operators) one obtains zero for the Casimir since $L^{MN}L_{MN}$ is
constructed from $X^{2},P^{2},X\cdot P$ which vanish in the physical sector.

\subsubsection{H-atom gauge}

Another solution of the constraints $X^{2}=P^{2}=X\cdot P=0,$ is
\cite{2tHandAdS}
\begin{align}
X^{M}  &  {=}{[}\overset{0^{\prime}}{r\cos u}{,\,~~}\overset{0}{r\sin u{\,}%
}{,~}\overset{1^{\prime}}{\frac{{\vec{r}\cdot\vec{p}}}{-{\alpha}}r\sqrt{-{2H}%
}}{,}\overset{i}{({\vec{r}}^{i}{-}\frac{{\vec{r}\cdot\vec{p}}}{{\alpha/r}%
}{\vec{p}}^{i}\mathbf{)}}\overset{~}{{]}}\;r{\equiv}\left\vert {\vec{r}%
}\right\vert \label{Hgauge}\\
P^{M}  &  {=}[\frac{-{\alpha}\sin{u}}{r}{\,\,,\,}\frac{{\,\alpha}\cos{u}}%
{r}{,\,}\left(  \frac{\alpha}{r}{-\vec{p}}^{2}\right)  {,\sqrt{-2H}~\vec{p}%
}^{i}{~~]\left(  -2H\right)  ^{-1/2}}\nonumber
\end{align}
where $H=\frac{\mathbf{\vec{p}}^{2}}{2}-\frac{\alpha}{r},\;$and$\;u=\left(
{\vec{r}\cdot\vec{p}-2\tau H}\right)  \frac{\sqrt{-{2H}}}{\alpha}.$ The
interpretation of the emergent dynamics is found by examining the gauge fixed
action
\[
\text{gauge fixed \ }S=\int d\tau\left(  {\partial}_{\tau}{\vec{r}\cdot\vec
{p}-H}\right)  \leftrightarrow\int d\tau(\frac{{1}}{{2}}\left(  \partial
_{\tau}{\vec{r}}\right)  ^{2}+\frac{\alpha}{{r}})
\]
Evidently, this is the spinless Hydrogen atom (or a planetary system, etc.).
The original SO$(d,2)$ global symmetry $L^{MN}=X^{M}P^{N}-X^{N}P^{M}$ must be
a hidden symmetry of this action. The gauge fixed generators are (at the
classical level)%
\begin{align*}
L^{0^{\prime}0}  &  =\frac{\alpha}{\sqrt{-2H}},\;\;L^{ij}=r^{i}p^{j}%
-r^{j}p^{i},\quad L^{1^{\prime}i}=\frac{1}{\sqrt{-2H}}\left(  {r\cdot p~p}%
^{i}{-r}^{i}{p}^{2}{-\alpha}\frac{\mathbf{r}^{i}}{r}\right)  ,\\
L^{0^{\prime}1^{\prime}}  &  =-r\cdot p\sin u+\frac{\alpha}{\sqrt{-2H}%
}(1-\frac{r\mathbf{p}^{2}}{\alpha})\cos u,\;\;L^{01^{\prime}}=r\cdot p\cos
u+\frac{\alpha}{\sqrt{-2H}}(1-\frac{r\mathbf{p}^{2}}{\alpha})\sin u,\\
L^{0^{\prime}i}  &  =rp^{i}\cos u~+\frac{\alpha}{\sqrt{-2H}}(\frac
{\mathbf{r}^{i}}{r}-\frac{\mathbf{r\cdot p}}{\alpha}p^{i})\sin u,\;~L^{0i}%
=rp^{i}\sin u~\,\,-\frac{\alpha}{\sqrt{-2H}}(\frac{\mathbf{r}^{i}}{r}%
-\frac{\mathbf{r\cdot p}}{\alpha}p^{i})\cos u
\end{align*}
In the first line one can recognize the angular momentum and the Runge-Lenz
vector that are long known to be conserved quantities of the H-atom system
(i.e. commute with $H$) and that they correspond to a hidden SO$\left(
d\right)  $ symmetry (better known as SU$\left(  2\right)  \times$SU$\left(
2\right)  $=SO$\left(  4\right)  $ in 3 space dimensions). However, 2T-physics
gives a stronger symmetry, not of the Hamiltonian, but of the action.
According to 2T-physics the H-atom action is invariant under SO$\left(
d,2\right)  .$ Indeed this symmetry can be verified directly\footnote{See a
homework problem and its solution at http://physics.usc.edu/\symbol{126}%
bars/papers/Hatom.pdf.}$.$ Before this was understood in 2T-physics no-one
seems to have been aware of the symmetry of the action, although there has
been discussions of a dynamical SO$\left(  4,2\right)  $ algebra of the H-atom system.

We can go further by quantizing the system, ordering properly the operators
and computing the Casimir operator. We find again that $C_{2}$ reduces to a
pure number which corresponds to the singleton representation \cite{2tHandAdS}%
\[
\text{{After\ quantum\ ordering}}{:\ }C_{2}=\frac{1}{2}L^{MN}L_{MN}%
=1-\frac{d^{2}}{4}%
\]
This is what it should be according to the general properties of 2T-physics.
Indeed $L_{MN}$ are gauge invariant and they should give the same Casimir in
any gauge. This also fits the idea of a duality between the free relativistic
particle and the H-atom, since these are derived from the same 2T action by
gauge fixing, and therefore they can be transformed to each other by the
Sp$\left(  2,R\right)  $ gauge transformations. Such Sp$\left(  2,R\right)  $
transformations are easily constructed classically between the gauges
(\ref{particlegauge}) and (\ref{Hgauge}). This is expected to succeed also at
the quantum level in the form of unitary transformations among dual bases,
since the quantum states in either gauge belong to the same representation of
SO$\left(  d,2\right)  $ with the same Casimir operators.

\subsubsection{More examples of emergent dynamics/spacetimes}

Many more 1T dynamical systems emerge holographically from the same 2T theory
given in Eq. (\ref{flat2T}). The diagram below illustrates some of the cases
that have been investigated \cite{2tHandAdS}.
\begin{center}
\fbox{\includegraphics[
height=3.4264in,
width=4.254in
]%
{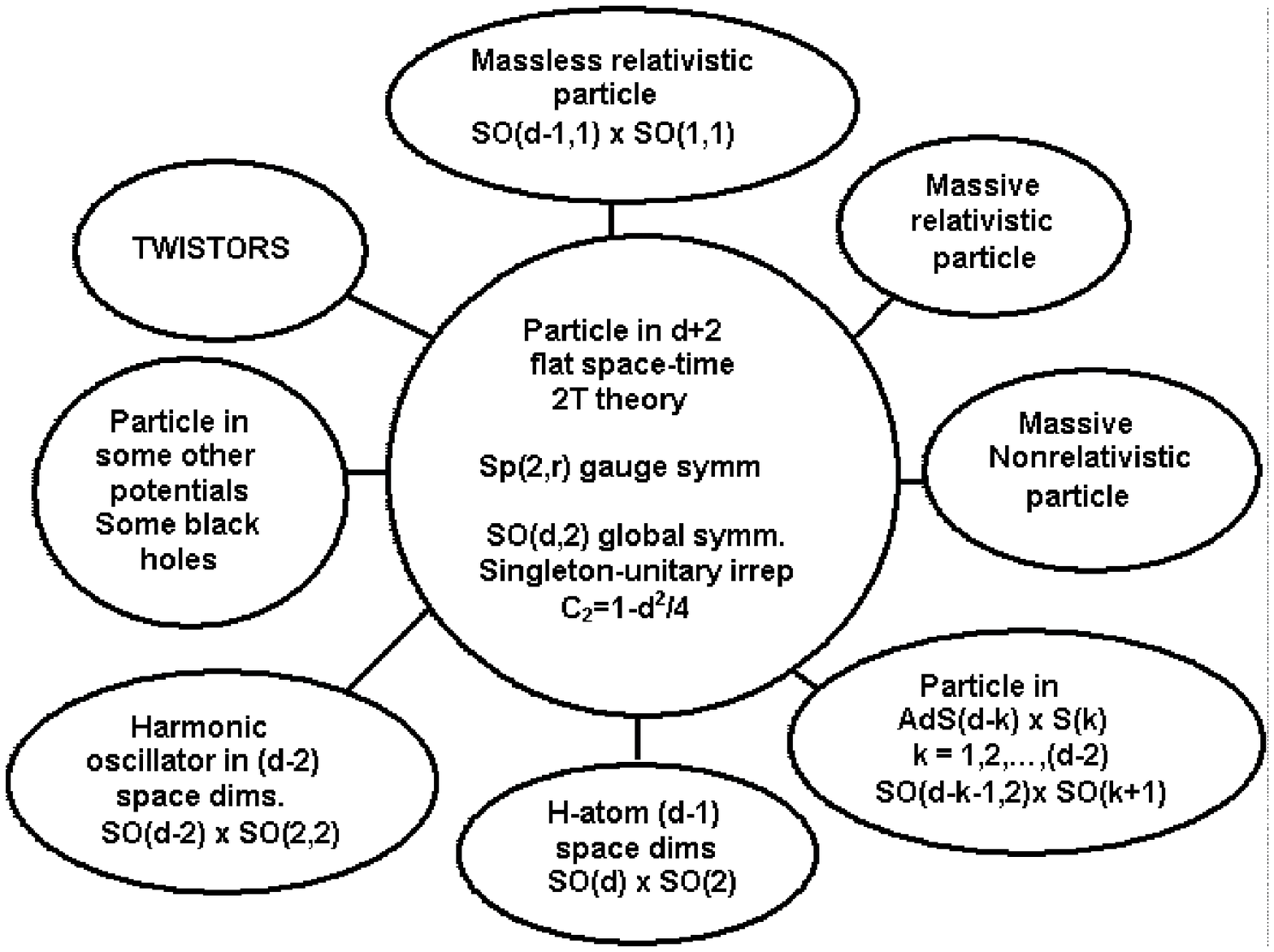}%
}\\
{\small d+2 to d holography gives many 1T systems.}\linebreak{\small Each one
is a basis for C}$_{2}${\small =1-d}$^{2}${\small /4 irrep of SO(d,2).}%
\label{corners}%
\end{center}
These include interacting as well as free systems. The quantum theory has been
investigated, and the quantum ordering of the $L^{MN}$ operators has been
obtained for the cases of the massless relativistic particle, H-atom, harmonic
oscillator, particle on AdS$_{d-k}\times$S$^{k}$ background, particle in the
SL$\left(  3,R\right)  $ black hole, and twistors. In each case it is shown
that $C_{2}=1-d^{2}/4.$

The 1T systems are derived from the same 2T system by making an Sp$\left(
2,R\right)  $ gauge choice. In each case some combination of $X^{M}\left(
\tau\right)  ,P^{M}\left(  \tau\right)  $ is gauge fixed to be $\tau.$ The
canonical conjugate to that choice is always the Hamiltonian written as a
function of the remaining phase space degrees of freedom. Although the
Hamiltonian (time) looks very different in each case, it still represents a
holographic image of the original 2T particle. Thus, each one of these systems
represents the same 2T theory although they each have a different
interpretation in 1T-physics. In the 1T context they must be interpreted as
being dual to each other. In the quantized version each one provides a basis
for the same singleton representation of SO$\left(  d,2\right)  .$ Within the
same representation they must correspond to different bases (which diagonalize
the respective Hamiltonian) related by unitary transformations. The existence
of such relationships are not at all evident in the 1T approach.

Further generalizations include 2T formulations of spinning particles
\cite{2tspinning}, spacetime supersymmetry \cite{2tSuper}, twistors
\cite{2ttwistor}\cite{2tsuperstring}\cite{twistorD}, and some study of
2T-physics in the context of field theory \cite{2tfield} and string theory
\cite{2tString}\cite{2tsuperstring}. A lot more basic research is awaiting to
be developed in 2T-physics.

With what we know so far about 2T-physics, it is evident that it applies
broadly to physics and correctly describes the physical world as we know it.
The advantage of 2T-physics over 1T-physics is its unification of various 1T
systems into a single 2T system, thus providing a more unified perspective.
This aspect could be illuminating for the physics that we already understand
in the 1T formalism by taking advantage of the revealed hidden symmetries and
by exploring the unsuspected duality type relationships among various 1T
dynamical systems. Knowing these facts should shed light on the solution and
interpretation of physical systems. Furthermore, the 2T approach could be used
as a tool for formulating the physics we don't fully understand yet, such as
M-theory. The mystery of M-theory has been my main motivation so far in
pursuing and developing this approach.

\section{Twistors as a gauge choice in 2T-physics}

Twistors were obtained as one of the possible gauge choices in 2T-physics in
\cite{2ttwistor}, and this has been further explored in \cite{2tsuperstring}
and \cite{twistorD}. For this purpose we consider a group or supergroup $G$
that contains SO$\left(  d,2\right)  $ as a subgroup ($G=$SO$\left(
d,2\right)  ~$is the smallest choice). A group element $g\left(  \tau\right)
\in G$ is introduced as a degree of freedom in addition to $X^{M}\left(
\tau\right)  ,P^{M}\left(  \tau\right)  ,$ and it is taken in the smallest
representation of $G$ such that SO$\left(  d,2\right)  \in G$ is the spinor
representation. When $G$ is a supergroup it contains spacetime fermionic
spinors that will be useful for spacetime supersymmetric theories. The group
element $g$ contains also more bosons beyond $\left(  X,P\right)  $ which, in
most cases, can be gauged away by additional gauge symmetries.

$g$ is taken as a singlet under the Sp$\left(  2,R\right)  $ gauge group,
while $\left(  X^{M},P^{M}\right)  $ form a doublet. Next we introduce a
further gauge symmetry embedded in $G$ that acts on the \textit{left side of
}$g,$\textit{ }as well as on $X^{M},P^{M}$ as specified below, to have the
\textit{correct number of physical degrees of freedom} for a (spinning)
particle or superparticle after gauge degrees of freedom are eliminated. On
the right side of $g$ we maintain a full global symmetry $G,$ therefore rows
of $g$ transform as spinors under the SO$\left(  d,2\right)  \in G.$ This is
where twistors in the spinor representation of SO$\left(  d,2\right)  $ will
come from.

In most applications $G$ is a supergroup, and the fermions in $g$ are used to
supersymmetrize the 2T system. However, it is also possible to discuss purely
bosonic cases and even specialize to $G=$Spin$\left(  d,2\right)  .$ In this
setting it is possible to choose gauges to eliminate degrees of freedom from
$g$ and/or from $X,P.$ If all of $g$ is eliminated, as in the purely bosonic
$G=$Spin$\left(  d,2\right)  $ case, we remain only with $X,P$ which give the
2T system discussed in the previous sections. However, if all of $X,P$ and
some of the $g$ is eliminated we remain with some of the degrees of freedom of
$g$ which describe the same 2T system in terms of twistors. If $G$ is an
appropriate supergroup that yields the superparticle in one gauge, then in the
twistor gauge (with $X,P$ completely eliminated) we obtain the
\textit{supertwistor} description of the superparticle. In this way we derive
the correct supertwistor representation of several systems of interest in
several dimensions as given in \cite{2ttwistor}\cite{2tsuperstring}%
\cite{twistorD} and briefly described here.

\subsection{Supersymmetric 2T-physics}

The most general case studied corresponds to the supersymmetrization of the
spinning particle of spin $n/2.$ For the spin=$n/2$ generalization we
introduce $n$ fermions ${\psi}_{1}^{M}{,\cdots,\psi}_{n}^{M}$ in addition to
$X^{M},P^{M},g.$ Although in this lecture we will mainly discuss the $n=0$
case (i.e. only $X,P,g$), we first give the more general Lagrangian%
\begin{gather}
L=\frac{{\eta}_{MN}}{{2}}\left(  {q}^{ab}{\partial}_{\tau}{Y}_{a}^{M}{Y}%
_{b}^{N}{-A}^{ab}{Y}_{\{a}^{M}{Y}_{b]}^{N}\right)  ~~\text{{ }}\underset{{
q}^{{ ab}}~=\text{metric}}{\text{{local OSp(n%
$\vert$%
2)}}}\nonumber\\
-\frac{{1}}{2^{[d/2]}}\left(  {L}^{MN}{+S}^{MN}\right)  {Str}\left(  {\Gamma
}^{{MN}}{\partial}_{\tau}{gg}^{-1}\right)  {~,~}\text{{ }}{g\in G}%
_{d}~\text{{supergroup}} \label{flatSuper}%
\end{gather}
For $n=0$ the first line is the same as Eq. (\ref{flat2T}) which describes the
scalar 2T particle. The second line generalizes it with spacetime
supersymmetry. In a specific gauge this system yields the standard
superparticle as one of the holographic images \cite{2tSuper}\cite{2ttwistor}.
The nonzero spin case with supersymmetry has not appeared yet in the
literature, it will be discussed in \cite{twistorD}, here we provide a brief description.

The first line in Eq. (\ref{flatSuper}) by itself describes the spinning
particle with spin $n/2$. This generalizes the scalar 2T-particle in Eq.
(\ref{flat2T}) by replacing the Sp$\left(  2\right)  $ doublet $\left(
X^{M},P^{M}\right)  $ by the OSp$\left(  n|2\right)  $ fundamental
representation with $n$ fermions, and introducing the corresponding gauge
fields ${A}_{a}^{b}$ (which include the Sp$\left(  2,R\right)  $ gauge fields
along with SO$\left(  d\right)  $ and fermionic counterparts)
\begin{equation}
{Y}_{a}{=}\underset{\text{fundamental of OSp(n%
$\vert$%
2)}}{\left(  {X}^{M}{,P}^{M}{,\psi}_{1}^{M}{,\cdots,\psi}_{n}^{M}\right)
},\;{D}_{\tau}{Y}_{a}^{M}=\underset{\text{OSp(n%
$\vert$%
2) gauge field}}{{\partial}_{\tau}{Y}_{a}^{M}-{A}_{a}^{b}{Y}_{b}^{M}%
,}\;\underset{S^{{MN}}={ \psi}_{i}^{{[M}}{ \psi}_{i}^{{N]}}~~~\text{{spin=}%
}\frac{{n}}{{2}}}{J^{{MN}}=q^{ab}Y_{{a}}^{{M}}Y_{{b}}^{{N}}=L^{{MN}}+S^{{MN}}%
},~ \label{ospn2}%
\end{equation}
This system (without the supersymmetrization of the second line) was discussed
in detail in \cite{2tspinning}. Since we don't have time to discuss it here,
we will specialize to only $n=0$ in this lecture.

The second line in Eq. (\ref{flatSuper}) corresponds to supersymmetrizing the
system in the first line with or without spin. Here the supergroup
$G_{\text{d}}$ with $N$ supersymmetries is taken for various dimensions $d$ as
one of the following supergroups
\[
G_{\text{d}}=\text{OSp}(N|4)_{\text{{3}}},~\text{SU}(2,2|N)_{\text{{4}}%
},~\text{F}(4)_{\text{{5}}},~\text{OSp}(8|N)_{\text{{6}}},~\text{PSU}%
(2,2|4)_{\text{{10}}}%
\]
When $d=3,4,5,6$ the spinor representation of SO$\left(  d,2\right)  $
corresponds to the first block in the fundamental matrix representation of
$G_{d}$ as shown below
\begin{equation}
\overset{\text{G}_{{d}}\text{=OSp(N%
$\vert$%
4)}_{{3}}\text{,SU(2,2%
$\vert$%
N)}_{{4}}\text{,F(4)}_{{5}}\text{,OSp(8%
$\vert$%
N)}_{{6}},\text{ etc.}}{\underset{d>6\text{ supergroups contain more than
Spin}\left(  d,2\right)  \text{ }\rightarrow\text{ }\Gamma^{M_{1}\cdots M_{p}%
}\text{D-branes}^{\text{\ref{Gbose}}}}{\text{{ g}}\left(  \tau\right)
=\exp\left(
\begin{tabular}
[c]{|l|l|}\hline
$\underset{\text{Spin}\left(  d,2\right)  ~\text{subgroup}}{\frac{1}{4}%
\Gamma^{MN}\omega_{MN}}$ & $\underset{\text{fermi}~}{\Theta_{{spinor}%
}^{{i=1\cdots N}}}$\\\hline
$\underset{\text{fermi}~}{\bar{\Theta}}$ & $\underset{~\text{R-symmetry}%
}{R^{a}\omega_{a}}$\\\hline
\end{tabular}
\ \right)  }} \label{group}%
\end{equation}
Therefore, in these cases the coupling $\left(  {L}^{MN}{+S}^{MN}\right)
{\Gamma}^{{MN}}{\ }$shown in Eq. (\ref{flatSuper}) takes the matrix form
$2^{-[d/2]}\left(  {L}^{MN}{+S}^{MN}\right)  \left(
\genfrac{}{}{0pt}{}{{\Gamma}^{{MN}}}{0}%
\genfrac{}{}{0pt}{}{0}{0}%
\right)  .$ This coupling scheme \cite{2tSuper}\cite{2ttwistor} applies to all
the cases listed in the first column of the table Eq.(\ref{table1}) and to the
SO$\left(  11,2\right)  $ covariant $d=11$ OSp$\left(  1|64\right)  $ toy
M-model \cite{2ttoyM} in the second column.

For the $d=10$ AdS$_{5}\times$S$^{5}$ case listed in the second column the
coupling scheme is slightly different. Namely, we do not keep full covariance
SO$\left(  10,2\right)  $ in 12 dimensions, but rather only under its subgroup
SO$\left(  4,2\right)  \times$SO$\left(  6\right)  $ = SU$\left(  2,2\right)
\times$SU$\left(  4\right)  .$ The supergroup that contains this subgroup is
PSU$\left(  2,2|4\right)  .$ To take this into account we split the 12
coordinates into two groups of six each $X^{M}=\left(  X^{m},X^{I}\right)  ,$
and similarly for the momenta $P^{M}=\left(  P^{m},P^{I}\right)  ,$ and
associate the first six dimensions with the upper block of the PSU$\left(
2,2|4\right)  $ matrix and the last six dimensions with the lower block (in
place of the R-symmetry). Therefore, for $d=10$ the coupling $\left(  {L}%
^{MN}{+S}^{MN}\right)  {\Gamma}^{{MN}}{\ }$shown in Eq. (\ref{flatSuper})
takes the matrix form $\frac{1}{4}\left(  {L}^{mn}{+S}^{mn}\right)  \left(
\genfrac{}{}{0pt}{}{{\Gamma}^{{mn}}}{0}%
\genfrac{}{}{0pt}{}{0}{0}%
\right)  +\frac{1}{4}\left(  {L}^{IJ}{+S}^{IJ}\right)  \left(
\genfrac{}{}{0pt}{}{{0}}{0}%
\genfrac{}{}{0pt}{}{0}{-{\Gamma}^{{IJ}}}%
\right)  .$ This $d=10$ 2T model, for $n=0$ (i.e. no spin $S^{MN}$ or $S^{IJ}%
$), produces the supersymmetric particle moving on AdS$_{5}\times$S$^{5}$ in
one of its holographic images, and its physical quantum spectrum is identical
to the Kaluza-Klein towers of type IIB supergravity compactified on
AdS$_{5}\times$S$^{5}$ \cite{2tAdSs}$.$ This is a model of interest in the
context of the $d=10$ twistor superstring described below, and the AdS-CFT
correspondence. A similar model for the d=11 AdS$_{4}\times$S$^{7}$ or
AdS$_{7}\times$S$^{4}$ listed in the table does not lead to the corresponding
compactification of $d=11$ supergravity, but rather to the first massive level
of the d=11 supermembrane.
\begin{equation}%
\begin{tabular}
[c]{|l|l|l|l|}\hline
$\text{d=3:}$ & $_{\text{Spin}\left(  3,2\right)  =\text{Sp}\left(  4\right)
\subset\text{OSp}(N|4)}$ & $\underset{\text{AdS}_{5}\times\text{S}^{5}%
}{\text{d=10}:}$ & $_{\text{Spin}\left(  4,2\right)  \times\text{Spin}\left(
6\right)  \subset\text{PSU}(2,2|4)}$\\\hline
$\text{d=4:}$ & $_{\text{Spin}\left(  4,2\right)  =\text{SU}\left(
2,2\right)  \subset\text{PSU}(2,2|N)}\;\;\;$ & $\underset{\text{AdS}_{4}%
\times\text{S}^{7}}{\text{d=11}:}$ & $_{\text{Spin}\left(  3,2\right)
\times\text{Spin}\left(  8\right)  \subset\text{OSp}(8|4)}$\\\hline
d=5 & $_{\text{Spin}\left(  5,2\right)  \subset~F\left(  4\right)  }~$\ $%
\genfrac{}{}{0pt}{}{\text{it contains}}{\text{also SU}\left(  2\right)  }%
$ & $\underset{\text{AdS}_{7}\times\text{S}^{4}}{\text{d=11}:}$ &
$_{\text{Spin}\left(  6,2\right)  \times\text{Spin}\left(  5\right)
\subset\text{OSp}(8|4)}$\\\hline
d=6 & $_{\text{Spin}\left(  6,2\right)  \subset\text{OSp}(8|N)}$ & d=11: &
$_{\text{Spin}\left(  11,2\right)  \subset\text{OSp}(1|64)}~%
\genfrac{}{}{0pt}{}{\text{toy~M-model}}{\text{with~D~branes}^{\ref{Gbose}}}%
$\\\hline
$\text{any d}$ & $_{\text{Spin}\left(  d,2\right)  =G}$~~$%
\genfrac{}{}{0pt}{}{\text{purely~bosonic}}{\text{twistors in any~d}}%
$ & etc. & {\small generalizations of above}\\\hline
\end{tabular}
\ \ \ \ \label{table1}%
\end{equation}
For $d>6$ supergroups contain bosonic subgroups that are larger than
SO$\left(  d,2\right)  $ as long as we insist that SO$\left(  d,2\right)  $
appears in the spinor representation\footnote{Here we give a list of the
smallest bosonic subgroups in a supergroup $G$ that contain Spin$\left(
d,2\right)  $ for $3\leq d\leq12.$ We also list the generators, and their
numbers in parentheses, as represented by antisymmetrized products of gamma
matrices $\Gamma^{M_{1}\cdots M_{n}}\equiv\frac{1}{n!}\Gamma^{\lbrack M_{1}%
}\cdots\Gamma^{M_{n}]}$ in dimension $d+2$ labelled my $M$ \label{Gbose}%
\[%
\begin{tabular}
[c]{|l|l|l|l|l|l|}\hline
d & Spin$\left(  d,2\right)  $ & spinor & $\subseteq$ G$_{bose}$ & generators
of G$_{bose}$ in Spin$\left(  d,2\right)  $ basis & contained in\\\hline
3 & Spin$\left(  3,2\right)  $ & 4 & Sp$\left(  4,R\right)  $ & $\Gamma^{MN}$
$\left(  10\right)  $ & $\left(  4\times4\right)  _{s}$\\\hline
4 & Spin$\left(  4,2\right)  $ & 4$_{\pm}$ & SU$\left(  2,2\right)  $ &
$\Gamma^{MN}$ $\left(  15\right)  $ & $\left(  4\times4^{\ast}\right)
$\\\hline
5 & Spin$\left(  5,2\right)  $ & 8$_{+}$ & SO$^{\ast}\left(  8\right)  $ &
$\Gamma^{MN}$ $\left(  21\right)  $ + $\Gamma^{M}$ $\left(  7\right)  $ &
$\left(  8\times8\right)  _{a}$\\\hline
6 & Spin$\left(  6,2\right)  $ & 8$_{+}$ & SO$^{\ast}\left(  8\right)  $ &
$\Gamma^{MN}$ $\left(  28\right)  $ & $\left(  8\times8\right)  _{a}$\\\hline
7 & Spin$\left(  7,2\right)  $ & 16 & SO$^{\ast}\left(  16\right)  $ &
$\Gamma^{MN}$ $\left(  36\right)  $ + $\Gamma^{MNK}$ $\left(  84\right)  $ &
$\left(  16\times16\right)  _{a}$\\\hline
8 & Spin$\left(  8,2\right)  $ & 16$_{\pm}$ & SU$^{\ast}\left(  16\right)  $ &
$\Gamma^{MN}$ $\left(  45\right)  $ + $\Gamma^{MNKL}$ $\left(  210\right)  $ &
$\left(  16\times16^{\ast}\right)  $\\\hline
9 & Spin$\left(  9,2\right)  $ & 32 & Sp$^{\ast}\left(  32\right)  $ &
$\Gamma^{MN}$ $\left(  55\right)  $ + $\Gamma^{M}$ $\left(  11\right)  $ +
$\Gamma^{M_{1}\cdots M_{5}}$ $\left(  462\right)  $ & $\left(  32\times
32\right)  _{s}$\\\hline
10 & Spin$\left(  10,2\right)  $ & 32$_{+}$ & Sp$^{\ast}\left(  32\right)  $ &
$\Gamma^{MN}$ $\left(  66\right)  $ + $\Gamma_{+}^{M_{1}\cdots M_{6}}$
$\left(  462\right)  $ & $\left(  32\times32\right)  _{s}$\\\hline
11 & Spin$\left(  11,2\right)  $ & 64 & Sp$^{\ast}\left(  64\right)  $ &
$\Gamma^{MN}$ $\left(  78\right)  $ + $\Gamma^{MNK}$ $\left(  286\right)  $ +
$\Gamma^{M_{1}\cdots M_{6}}$ $\left(  1716\right)  $ & $\left(  64\times
64\right)  _{s}$\\\hline
12 & Spin$\left(  12,2\right)  $ & 64$_{\pm}$ & SU$^{\ast}\left(  64\right)  $
& $\Gamma^{MN}$ $\left(  91\right)  $ + $\Gamma^{MNKL}$ $\left(  1001\right)
$ + $\Gamma^{M_{1}\cdots M_{6}}$ $\left(  3003\right)  $ & $\left(
64\times64^{\ast}\right)  $\\\hline
\end{tabular}
\ \ \ \
\]
The antisymmetric $\Gamma^{M_{1}\cdots M_{n}}$ are associated with group
parameters $\omega_{M_{1}\cdots M_{n}}\left(  \tau\right)  $ that cannot be
eliminated by the gauge symmetries, and therefore they are additional degrees
of freedom analogous to D-brane collective coordinates.}. The extra bosons
contained in $g$ correspond to D brane-like degrees of freedom. If we require
full covariance for SO$\left(  d,2\right)  $ we are forced to admit these as
additional degrees of freedom beyond those of the superparticle. The toy
M-model in $d=11$ based on OSp$\left(  1|64\right)  $ is one of the most
interesting cases of this type\cite{2ttoyM}. By breaking the covariance to a
subgroup of SO$\left(  d,2\right)  $ we can build 2T models such as the
AdS$_{d-k}\times$S$^{k}$ cases based only on particle degrees of freedom
similar to what was described above.

\subsection{Twistor gauge}

The Lagrangian in Eq. (\ref{flatSuper}) has an evident global symmetry $G_{d}$
which corresponds to group transformations on the right side of $g.$ These
leave the Cartan form $\left(  \partial g\right)  g^{-1}$ invariant. The
conserved Noether charge for this symmetry is the supermatrix ${J}_{{A}}%
^{~{B}}$ in the Lie algebra of $G_{d}$
\[
\text{{ Gauge invariant global symm}}\;\text{{ \ }}{J}_{{A}}^{~{B}}\sim
\frac{i}{2}\left(  {L}_{{MN}}+S_{MN}\right)  \left(  {g}^{-1}{\Gamma}^{{MN}%
}{g}\right)  _{{A}}^{~{B}}%
\]
The Lagrangian is also invariant under a number of local symmetries. To begin
there is the built in OSp$\left(  n|2\right)  $ local supersymmetry on the
worldline as in Eq. (\ref{ospn2}). In addition, there are local spacetime
supersymmetries. Some of these become easier to notice by rewriting the
Lagrangian (\ref{flatSuper}) in the form%
\[
L=\frac{{1}}{{2}^{\left[  {d/2}\right]  }}q^{ab}Tr\left[  \partial_{\tau
}\left(  g^{-1}Y_{a}\cdot\Gamma g\right)  \left(  g^{-1}Y_{b}\cdot\Gamma
g\right)  \right]  -\frac{{1}}{{2}}A^{ab}Y_{a}\cdot Y_{b}%
\]
Then it is easy to see that there is an invariance under local transformations
[Spin$\left(  d,2\right)  $ $\times$ R-symmetry]$\in G_{d}$ that are
simultaneously applied on the \textit{left side} of $g$ as well as on the $M$
index of $Y_{a}^{M}.$ There is also local fermionic kappa supersymmetry that
is also applied on the left side of $G_{d}$ as well as on $A^{ab}$ as
explained in \cite{2tSuper}\cite{2ttwistor}\cite{2tsuperstring}. Let us review
what physical degrees of freedom remain after gauge degrees of freedom are
removed. We consider $n=0$ in what follows (i.e. supersymmetrizing the scalar
particle), the general $n$ is similar.

The local Spin$\left(  d,2\right)  $ $\times$ R-symmetry has enough gauge
parameters to remove all of the bosons from $g$ in all the cases listed in the
table Eq.(\ref{table1}), except for the toy M-model which has
D-branes$^{\ref{Gbose}}$. Therefore for those cases the bosonic degrees of
freedom are just the particle phase space $\left(  X^{M},P^{M}\right)  $ or
their gauge equivalent.

The local kappa supersymmetry has enough fermionic gauge parameters to remove
$3/4$ of the fermions $\Theta$ shown in Eq.(\ref{group}) for $d=3,4,5,6$ and
the $d=11$ toy M-model. On the other hand for the $d=10$ AdS$_{5}\times$%
S$^{5}$ case only $1/2$ of the 32 fermions can be removed by the kappa
supersymmetry, while for the $d=11$ AdS$_{4}\times$S$^{7}$ or AdS$_{7}\times
$S$^{4}$ there is no kappa supersymmety at all.

This summarizes then the physical degrees of freedom up to gauge equivalences.
The interesting aspect of 2T-physics is that the gauge equivalence within the
2T-system does not necessarily imply that the 1T interpretation is the same,
but rather that there are dualities between various holographic 1T images, as
in the figure above.

It was shown in \cite{2tSuper}\cite{2ttwistor}\cite{2tsuperstring} that, for
$n=0$ and $d=3,4,5,6$, one can choose a gauge that reduces the 2T system to
the standard superparticle in the corresponding number of dimensions. Also it
was shown in \cite{2tAdSs} that the $d=10$ SU$\left(  2,2|4\right)  $ case can
be gauge fixed to the superparticle moving on the AdS$_{5}\times$S$^{5}$ background.

In this lecture we will concentrate on the twistor gauge \cite{2ttwistor}%
\cite{2tsuperstring} for the $n=0$ case. By using the local symmetry
Sp$\left(  2,R\right)  \times$Spin$\left(  d,2\right)  ,$ and the constraints
$X\cdot P=X^{2}=P^{2}=0$, we can gauge fix $X^{M}\left(  \tau\right)
,P^{M}\left(  \tau\right)  $ for all $\tau$ to the following trivial
configuration
\begin{equation}
{X}^{M}{=}(\overset{+^{\prime}}{{X}^{+^{\prime}}},\overset{-^{\prime}}{{0}%
},\overset{+}{{0}},\overset{-}{{0}},\overset{i}{{0}}),\;{P}^{M}{=}%
(\overset{+^{\prime}}{{0}},\overset{-^{\prime}}{{0}},\overset{+}{{P}^{+}%
},\overset{-}{{0}},\overset{i}{{0}}),\;i=1,\cdots,\left(  d-2\right)
.\label{twistgauge1}%
\end{equation}
In this gauge the purely bosonic system for any $d,$ the supersymmetric
systems for $d=3,4,6,$ and the $d=11$ toy M-model listed in Eq. (\ref{table1})
collapse to the form\footnote{For the $d=10$ AdS$_{5}\times$S$^{5}$ case the
gauge fixed forms of $X,P$ are different as given in Eq.(\ref{asdgauge}).}%
\begin{align}
L &  =-\frac{L^{+^{\prime}+}}{{2}^{[{d/2]-1}}}Tr\left(  \partial_{\tau}%
gg^{-1}\left(
\genfrac{}{}{0pt}{}{{\Gamma}^{{-}^{\prime}-}}{0}%
\genfrac{}{}{0pt}{}{0}{0}%
\right)  \right)  =-i\partial_{\tau}\bar{Z}^{aA}Z_{Aa}\;,\;a=1,\cdots
,k\label{twistL}\\
J_{{A}}^{~{B}} &  \sim L^{+^{\prime}+}\left(  g^{-1}\left(
\genfrac{}{}{0pt}{}{{\Gamma}^{{-}^{\prime}-}}{0}%
\genfrac{}{}{0pt}{}{0}{0}%
\right)  g\right)  _{{A}}^{~{B}}=-2Z_{Aa}\bar{Z}^{aB}\;\;,\;\bar{Z}^{aA}%
Z_{Ab}=0\label{twistJ}%
\end{align}
In an appropriate basis for gamma matrices\footnote{$d+2$ gamma matrices that
satisfy $\Gamma^{M}\bar{\Gamma}^{N}+\Gamma^{N}\bar{\Gamma}^{M}=2\eta^{MN}$ are
chosen in a Weyl basis as follows%
\[
\Gamma^{\pm^{\prime}}=\left(  \pm\sqrt{2}\tau^{\pm}\right)  \otimes
1\otimes1_{k},\;\;\;\Gamma^{\pm}=\tau_{3}\otimes\left(  \pm\sqrt{2}\sigma
^{\pm}\right)  \otimes1_{k},\;\;\;\Gamma_{i}=\tau_{3}\otimes\sigma_{3}%
\otimes\gamma_{i},\;%
\genfrac{}{}{0pt}{}{i=1,\cdots,\left(  d-2\right)  }{\gamma_{i}~\text{is
k}\times\text{k matrix}\;}%
\;
\]
In odd dimensions we have $\bar{\Gamma}^{M}=\Gamma^{M}$. In even dimensions
$\bar{\Gamma}^{M}$ differs from $\Gamma^{M}$ only for the last gamma matrix,
which is proportional to the identity $1_{4k}=1_{2}\otimes1_{2}\otimes1_{k}$
$,$ namely $\Gamma_{d-2}=i1_{4k}=-\bar{\Gamma}_{d-2}$. For example for $d=4$
or SO$\left(  4,2\right)  =$SU$\left(  2,2\right)  ,$ we have $k=1,$ we choose
$\gamma_{1}=-1,$ and take $\Gamma_{2}=i1_{4}=-\bar{\Gamma}_{2}.$ Then we
construct $\Gamma^{MN}=\frac{1}{2}\left(  \Gamma^{M}\bar{\Gamma}^{N}%
-\Gamma^{N}\bar{\Gamma}^{M}\right)  $ as $4\times4$ matrices.} we find
$\Gamma^{{-}^{\prime}-}=2\tau^{-}\otimes\sigma^{-}\otimes1_{k}$ which is a
$4k\times4k$ matrix with lots of zeroes and $k$ nonzeroes off the diagonal.
Therefore only certain off-diagonal rows of $g$ denoted by $\bar{Z}^{aA}$ and
certain off-diagonal columns of $g^{-1}$ denoted by $Z_{Aa}$ contribute in the
trace in $L$ or to $J_{{A}}^{~{B}}$. Also the relation $gg^{-1}=1$ implies the
constraint $\bar{Z}^{aA}Z_{Ab}=0$ as an off diagonal entry in the matrix $1.$
The $A,B$ indices label the fundamental representation of $G_{d}$ and
therefore the $Z_{Aa}$ denote $k$ supertwistors with $a=1,\cdots,k$. Thus the
theory has now been written in terms of twistors.

Note that for $d=4$ the group is PSU$\left(  2,2|4\right)  ,$ the gamma
matrices are $4\times4,$ and $k=1.$ Therefore for $d=4$ there is a single
supertwistor $Z_{A}$ in the fundamental representation of PSU$\left(
2,2|4\right)  $ and it is constrained by $\bar{Z}^{A}Z_{A}=0.$ These
constrained twistors describe CP$^{3|4}.$ Thus the 2T formalism for
supertwistors is in full agreement with the expectation about twistors in four
dimensions. The 2T formalism gives the appropriate generalization to all the
other dimensions mentioned earlier. These will be described below case by case
for a few dimensions of special interest.

The Lagrangian in (\ref{twistL}) suggests that $\bar{Z}^{aA}$ is the canonical
conjugate to $Z_{Aa}$ and therefore the twistors can be expressed in terms of
oscillators. The current $J_{A}^{~B}$ for the global symmetry in
Eq.(\ref{twistJ}) is constructed from these oscillators, and the quantum
states are obtained in the Fock space of these oscillators. The physical
states are the subset of the Fock space that satisfies the constraint $\bar
{Z}^{aA}Z_{Ab}=0,$ and form a unitary representation of the global symmetry
$G_{d}.$ This setup precisely coincides with the Bars-G\"{u}naydin (BG)
oscillator approach to unitary representations of supergroups developed in
1983 \cite{barsgunaydin}. The additional constraint is a gauge invariance
condition and is implemented by following the discussion about
\textquotedblleft color" in the improved oscillator formalism given in
\cite{2tZero}. Therefore, we can easily obtain the quantum spectrum and
compare to the quantum spectrum in another gauge, such as the superparticle
gauge. The agreement is perfect as expected from the 2T approach, since each
gauge corresponds to a holographic image of the same 2T system.

The supertwistors are constrained as shown above. The full solution of these
constraints in terms of unconstrained degrees of freedom is given as a coset
$T_{\Gamma}\in G_{d}/H_{\Gamma},$ where $H_{\Gamma}$ is a gauged subgroup
$H_{\Gamma}$ of $G_{d},$ that is a remainder of the original gauge symmetries
mentioned before. $H_{\Gamma}$ is identified as the subgroup that contains all
the generators of $G_{d}$ that commute with the generator represented by
$\left(
\genfrac{}{}{0pt}{}{{\Gamma}^{{-}^{\prime}-}}{0}%
\genfrac{}{}{0pt}{}{0}{0}%
\right)  .$ We can then show \cite{2tsuperstring}\cite{twistorD} that the Lie
algebras of $h_{\Gamma}$ and of the coset $t_{\Gamma}$ form~triangular
sub-supergroups, and they satisfy (anti)commutation rules of the type
\cite{2tsuperstring}\cite{twistorD}
\begin{equation}
\lbrack h_{\Gamma},h_{\Gamma}\}\sim h_{\Gamma},\;\;[t_{\Gamma},t_{\Gamma
}\}\sim t_{\Gamma},\;\;[h_{\Gamma},t_{\Gamma}\}\sim h_{\Gamma}+t_{\Gamma}
\label{ht1}%
\end{equation}
Furthermore the system can be written in terms of the unconstrained degrees of
freedom $t\in G_{d}/H_{\Gamma}$ in the form
\begin{equation}
L=-\frac{L^{+^{\prime}+}}{{2}^{[{d/2]-1}}}Tr\left(  \partial_{\tau}%
tt^{-1}\left(
\genfrac{}{}{0pt}{}{{\Gamma}^{{-}^{\prime}-}}{0}%
\genfrac{}{}{0pt}{}{0}{0}%
\right)  \right)  ,\;\;J_{{A}}^{~{B}}\sim L^{+^{\prime}+}\left(  t^{-1}\left(
%
\genfrac{}{}{0pt}{}{{\Gamma}^{{-}^{\prime}-}}{0}%
\genfrac{}{}{0pt}{}{0}{0}%
\right)  t\right)  _{{A}}^{~{B}} \label{ht2}%
\end{equation}
This is like a sigma model based on a coset but the Lagrangian is linear in
the Cartan connection $\partial_{\tau}tt^{-1}$ (as opposed to quadratic form
for the sigma model) and there is an unusual insertion $\left(
\genfrac{}{}{0pt}{}{{\Gamma}^{{-}^{\prime}-}}{0}%
\genfrac{}{}{0pt}{}{0}{0}%
\right)  .$

\subsection{Supertwistors for d=4 and SYM spectrum}

The twistor must reproduce the physical degrees of freedom and quantum states
of the corresponding $d=4,$ $N=4$ superparticle, as expected from the 2T
formalism. Let's see how this is obtained explicitly.

To begin the superparticle has $4x$,$4p$ and 16$\theta$ real degrees of
freedom in super phase space. We remove $1x$ and $1p,$ due to $\tau$
reparametrization and the corresponding $p^{2}=0$ constraint. We also remove
$8$ fermionic degrees of freedom due to kappa supersymmetry. We are left over
with $3x,3p,8\theta$ physical degrees of freedom. With these we construct the
physical quantum states as an arbitrary linear combination of the basis states
in momentum space $|\vec{p},\alpha\rangle,$ where $\alpha$ is the basis for
the Clifford algebra satisfied by the $8\theta.$ This basis has 8 bosonic
states and 8 fermionic states. Viewed as probability amplitudes in position
space $\langle x,\alpha|\psi\rangle$ these are equivalent to fields
$\psi\left(  x\right)  _{8_{B}+8_{F}}$ which correspond to the independent
\textit{solutions} of all the constraints. One finds that these are the same
as the 8 bose and 8 fermi fields of the Super Yang Mills (SYM) theory which
are the solutions of the linearized equations of motion in the lightcone
gauge. They consist of two helicities of the gauge field $A_{\pm1}\left(
x\right)  ,$ two helicities for the gauginos $\psi_{+\frac{1}{2}}^{a}\left(
x\right)  ,$ $\bar{\psi}_{-\frac{1}{2},a}\left(  x\right)  $ in the
$\mathbf{4,\bar{4}}$ of SU$\left(  4\right)  ,$ and six scalars $\phi^{\lbrack
ab]}\left(  x\right)  $ in the $\mathbf{6}$ of SU$\left(  4\right)  .$

Now we count the physical degrees of freedom for the twistors. We have already
explained following Eq. (\ref{twistL}) that for $d=4$ we have one complex
twistor $Z_{A}$ in the fundamental representation of PSU$\left(  2,2|4\right)
,$ with a Lagrangian and a conserved current given by
\begin{equation}
\underset{Z_{A}\text{{ \ is in fundamental representation of PSU(2,2%
$\vert$%
N)~}}\leftrightarrow\text{CP}^{3|N}}{L=i\bar{Z}^{A}\partial_{\tau}%
Z_{A},\;J_{A}^{~B}=-2Z_{A}\bar{Z}^{B},\;\text{and\ }\bar{Z}^{A}Z_{A}=0}
\label{4Dtwist}%
\end{equation}
To start $Z_{A}$ has $4$ complex bosons and 4 complex fermions, i.e.
$8_{B}+8_{F}$ real degrees of freedom. However, there is one constraint
$\bar{Z}^{A}Z_{A}=0$ and a corresponding U$\left(  1\right)  $ gauge
symmetry\footnote{This can be restated by reformulating the above system by
rewriting $L=\bar{Z}^{A}\left(  \partial_{\tau}+A\right)  Z_{A}$ with an extra
U$\left(  1\right)  $ gauge field $A$, and deriving the constraint by varying
$A$.}, which remove 2 bosonic degrees of freedom. The result is $6_{B}+8_{F}$
physical degrees of freedom which is equivalent to CP$^{3|4}$, and the same
number as $3x,3p,8\theta$ for the superparticle, as expected. Instead of
constrained twistors we can also express the $CP^{3|4}$ theory in terms of
unconstrained coset parameters in the form (\ref{ht1},\ref{ht2}) where
$h_{\Gamma}$ was given in \cite{2tsuperstring}, and with more details in
\cite{twistorD}.

To construct the spectrum we could resort to well known twistor techniques by
working with fields $\phi\left(  Z\right)  $ that are holomorphic in $Z_{A}$
on which $\bar{Z}^{A}$ acts as a derivative $\bar{Z}^{A}\phi\left(  Z\right)
=\partial\phi\left(  Z\right)  /\partial Z_{A},$ as dictated by the canonical
structure that follows from the Lagrangian (\ref{4Dtwist}). Imposing the
constraint amounts to requiring $\phi\left(  Z\right)  $ to be homogeneous
with a given degree $h$, namely $Z_{A}\bar{Z}^{A}\phi\left(  Z\right)
=Z_{A}\partial\phi\left(  Z\right)  /\partial Z_{A}=h\phi\left(  Z\right)  .$
Only one value of $h$ is permitted. Naively $h$ is zero at the classical
level, but at the quantum level we have to determine the correct value of $h$
that may arise due to quantum ordering. In the case of the $d=4$ $N=4$
superparticle described by the PSU$\left(  2,2|4\right)  $ twistor indeed we
find $h=0,$ and the resulting spectrum is again the SYM fields. This is the
degree zero wavefunction $\phi\left(  Z\right)  $ described in \cite{berkWit}.
Recall that in \cite{berkWit} there are also twistor wavefunctions $f\left(
Z\right)  ,g\left(  Z\right)  $ that describe the spectrum of conformal
gravity; those arise in the same twistor formalism, but at a different value
of $h.$ However, since only one value of $h$ is permitted in the current
superparticle model, only SYM states $\phi\left(  Z\right)  $ are present. Of
course, this is no surprise in the 2T setting. We have simply compared two
gauges, and we must agree.

It is also worth analyzing the quantum system in terms of oscillators related
to twistors and understand the unitarity of the physical space. We emphasize
that $\bar{Z}^{A}$ is obtained from $Z_{A}$ by hermitian conjugation and
multiplying by the PSU$\left(  2,2|4\right)  $ metric. To see the oscillator
formalism clearly we work in a basis of SU$\left(  2,2|4\right)  $ in which
the metric is $diag\left(  1_{2},-1_{2},1_{4}\right)  .$ This is the
SU$\left(  2\right)  \times$SU$\left(  2\right)  $ basis for SU$\left(
2,2\right)  ,$ to be contrasted with the SL$\left(  2,C\right)  $ basis in
which the metric is off-diagonal and usually used to discuss Lorentz covariant
twistors. The two bases are simply related by a linear transformation that
diagonalizes the metric. In this diagonal basis we identify the oscillators as
(a bar over the symbol means hermitian conjugation)%
\begin{align*}
L  &  =i\bar{Z}^{A}\partial_{\tau}Z_{A}=i\bar{a}^{i}\partial_{\tau}%
a_{i}-ib_{I}\partial_{\tau}\bar{b}^{I}+i\bar{\psi}^{r}\partial_{\tau}\psi
_{r}\;\;\;\;\;%
\genfrac{}{}{0pt}{}{\overset{\text{SU}\left(  2\right)  }{i=1,2}%
,\;\overset{\text{SU}\left(  2\right)  }{I=1,2}\;}{r=1,\cdots,4\;\text{SU}%
\left(  4\right)  }%
\\
Z_{A}  &  =\left(
\begin{array}
[c]{c}%
a_{i}\\
\bar{b}^{I}\\
\psi_{r}%
\end{array}
\right)  ,\;\bar{Z}^{A}=\left(  \bar{a}^{j},-b_{J},\bar{\psi}^{s}\right)
,\;\;J_{A}^{B}=-2Z_{A}\bar{Z}^{A}=-2\left(
\begin{array}
[c]{ccc}%
a_{i}\bar{a}^{j} & -a_{i}b_{J} & a_{i}\bar{\psi}^{s}\\
\bar{b}^{I}\bar{a}^{j} & -\bar{b}^{I}b_{J} & \bar{b}^{I}\bar{\psi}^{s}\\
\psi_{r}\bar{a}^{j} & -\psi_{r}b_{J} & \psi_{r}\bar{\psi}^{s}%
\end{array}
\right)
\end{align*}
It is significant to note that, after taking care of the metric in $\bar{Z}$
as above, the usual canonical rules applied to this Lagrangian identifies the
oscillators as being all \textit{positive norm} oscillators $\left[
a_{i},\bar{a}^{j}\right]  =\delta_{i}^{j},$ $\left[  b_{I},\bar{b}^{J}\right]
=\delta_{I}^{J}$ and $\left\{  \psi_{r},\bar{\psi}^{s}\right\}  =\delta
_{r}^{s}$ . Therefore all Fock states have positive norm. However, among them
we must choose only those that satisfy the constraints
\begin{align*}
0  &  =\bar{Z}^{A}Z_{A}=\bar{a}^{i}a_{i}-b_{I}\bar{b}^{I}+\bar{\psi}^{r}%
\psi_{r}=\bar{a}^{i}a_{i}-\left(  \bar{b}^{I}b_{I}+2\right)  +\bar{\psi}%
^{r}\psi_{r}\\
&  \Leftrightarrow2=N_{a}+N_{\psi}-N_{b}\equiv\Delta,\;\;\;N_{a},N_{\psi
},N_{b}~\text{number operators}%
\end{align*}
This is precisely the BG oscillator formalism for unitary representations of
noncompact groups \cite{barsgunaydin} for a single \textquotedblleft color".
The constraint $\Delta=2$ is discussed in \cite{2tZero}. These physical states
are characterised by identifying the following lowest supermultiplet%
\[
\Delta=2:\left(  \underset{\left(  1,0,1\right)  }{\overset{A_{+1}}{{\bar{a}%
}^{i}{\bar{a}}^{j}}}{,~}\underset{(\frac{1}{2},0,4)}{\overset{\psi_{+1/2}^{r}%
}{{\bar{a}}^{i}{\bar{\psi}}^{r}}}{,~}\underset{(0,0,6)}{\overset{\phi^{\left[
rs\right]  }}{{\bar{\psi}}^{r}{\bar{\psi}}^{s}}}{,~}\underset{(0,\frac{1}%
{2},\bar{4})}{\overset{\psi_{-1/2,a}}{{\bar{b}}^{I}{\bar{\psi}}^{r}{\bar{\psi
}}^{s}{\bar{\psi}}^{m}}}{,~}\underset{(0,1,1)}{\overset{A_{-1}}{{\bar{b}}%
^{I}{\bar{b}}^{J}{\bar{\psi}}^{r}{\bar{\psi}}^{s}{\bar{\psi}}^{m}{\bar{\psi}%
}^{n}}}\right)  |0\rangle\
\]
which is annihilated by the double annihilation generators $a_{i}b_{J}$ which
is part of $J_{A}^{B}$ in the conformal subgroup SU$\left(  2,2\right)  .$ All
other states with $\Delta=2$ are descendants obtained by applying arbitrary
powers of the double creation generator $\bar{a}^{j}\bar{b}^{I}$ in SU$\left(
2,2\right)  .$ The lowest states correspond to the SYM fields, the descendants
are analogous to applying multiple derivatives on a field. The classification
of the lowest states under SU$\left(  2\right)  \times$SU$\left(  2\right)
\times$SU$\left(  4\right)  \subset$PSU$\left(  2,2|4\right)  $ is given under
each combination of oscillators in the form $\left(  j_{1},j_{2},\dim\left(
SU\left(  4\right)  \right)  \right)  $ where $j_{1},j_{2}$ correspond to the
spin quantum numbers in each SU$\left(  2\right)  $. In arriving at these
quantum numbers we took into account that ${\bar{a}}^{i}{\bar{a}}^{j}$ is
symmetric while ${\bar{\psi}}^{r}{\bar{\psi}}^{s}$ is antisymmetric, etc.
Above each of the oscillator combination we indicated one of the physical
helicity components of the SYM fields associated with that state. This is
because in comparing the compact SU$\left(  2\right)  \times$SU$\left(
2\right)  \subset$SU$\left(  2,2\right)  $ to the \textit{helicity} embedded
in the noncompact Lorentz group SL$\left(  2,C\right)  \subset$SU$\left(
2,2\right)  $ we must identify as helicity only the spin up part from the
first SU$\left(  2\right)  $ and the spin down part from the second SU$\left(
2\right)  .$

Although we gave a whole supermultiplet of lowest states above, there really
is only one lowest oscillator state for the entire unitary representation of
PSU$\left(  2,2|4\right)  $. That one is simply ${\bar{\psi}}^{r}{\bar{\psi}%
}^{s}|0\rangle,$ which satisfies $\Delta=2.$ All other states with $\Delta=2$
are obtained by applying all powers of $J_{A}^{B}$ on this state (note
$[\Delta,J_{A}^{~B}]=0$). This is called the doubleton representation of
PSU$\left(  2,2|4\right)  $ (sometimes also called the singleton). If we watch
carefully the orders of the oscillators we can show that the generators of
PSU$\left(  2,2|4\right)  $ \textit{in this representation} satisfy
\cite{2tZero} the following nonlinear constraints%
\begin{equation}
\left(  JJ\right)  _{A}^{~B}=4\left(  J\right)  _{A}^{~B}+0 \label{JJ}%
\end{equation}
The linear $J$ follows from the commutation rules among the generators, the
coefficient $4$ is related to overall normalization of $J,$ while the
coefficient $0$ is the PSU$\left(  2,2|4\right)  $ quadratic Casimir
eigenvalue $C_{2}=0$. This equation should be viewed as a set of constraints
that are satisfied by the generators in this particular representation, and as
such this relation identifies uniquely the representation (there is a unique
$C_{2}=0$ representation if we also specify the SU$\left(  2,2\right)  $
conformal dimension=1). If the theory is expressed in any other form (such as
particle description, or field theory) the doubleton representation can be
identified in terms of the global symmetry as one that must satisfy the
constraints (\ref{JJ}), automatically requiring the 6 scalars $\phi^{\left[
ab\right]  }$ as the lowest SU$\left(  4\right)  $ multiplet. This is a
completely PSU$\left(  2,2|4\right)  $ covariant and gauge invariant way of
identifying the physical and unitary states of the theory. Of course the
$d=4,$ $N=4$ SYM fields satisfy this criterion.

\subsection{Supertwistors for d=6 and self dual supermultiplet}

Now that the concepts have been illustrated clearly for $d=4,$ we summarize
quickly the equivalent statements for $d=6.$ The superparticle in $d=6$ and
$N=4$ starts out with $6x,6p,16\theta$ real degrees of freedom. Fixing $\tau,$
and kappa local gauges and solving constraints, reduces the physical degrees
of freedom down to 5$x$, 5$p$, 8 $\theta.$ The superparticle action has a
hidden global superconformal symmetry OSp$(8^{\ast}|4)$ \cite{2tSuper}%
\cite{2ttwistor}, therefore the physical states should be classified as a
unitary representation under this group.

If we quantize in the lightcone gauge we find $8_{B}+8_{F}$ states which
should be compared to the fields of a six dimensional field theory taken in
the lightcone gauge. There are two possible candidates; (1) the linearized six
dimensional SYM theory with $N=4$ SUSY in the lightcone gauge, and (2) the
self dual supermultiplet classified (covariantly) as
\begin{equation}
\underset{\text{self dual }F_{[\mu\nu\lambda]}^{+}=\partial_{\lbrack\mu_{1}%
}A_{\mu_{2}\mu_{3}]}=\varepsilon_{\mu_{1}\mu_{2}\mu_{3}\mu_{4}\mu_{5}\mu_{6}%
}\partial^{\lbrack\mu_{4}}A^{\mu_{5}\mu_{6}]}}{\text{{ SO(5,1)}}%
\times\text{{Sp(4):}}\;F_{[\mu\nu\lambda]}^{+},\;\psi_{\alpha}^{a}%
,\;\phi^{\lbrack ab]}} \label{6Dfields}%
\end{equation}
The SYM lightcone degrees of freedom consists of 8$_{B}+8_{F},$ with the 8
bosons being: the transverse 4-vector $A_{i}$ in SO$\left(  4\right)  \subset
$SO$(5,1)$ and four real scalars $\phi^{I}$ in an internal compact SO$\left(
4\right)  $. On the other hand for the self dual multiplet, we have the
following lightcone fields: a self dual antisymmetric tensor $A_{ij}%
=i\varepsilon_{ijkl}A^{kl}$ in SO$\left(  4\right)  \subset$SO$(5,1)$ (i.e. 3
fields), and the Sp$\left(  4\right)  $ traceless antisymmetric $\phi^{\lbrack
ab]}$ (5 scalars). These are clearly different. Only the self dual
supermultiplet is consistent with the compact USp$\left(  4\right)  \subset
$OSp$(8^{\ast}|4)$ classification (the fundamental $\mathbf{4}$ of USp$\left(
4\right)  $ is not real but pseudo-real, while the $\mathbf{5}$ represented as
$\phi^{\lbrack ab]}$ is real). Therefore the initial hidden superconformal
symmetry OSp$(8^{\ast}|4)$ of the superparticle (and of the 2T superparticle)
is consistent only with the field theory for the $d=6$ self dual
supermultiplet. An \textit{interacting} quantum conformal field theory with
these degrees of freedom is expected but cannot be written down covariantly in
the form of a field theory \cite{witten6D}.

Let us now examine the twistors that emerged in Eq. (\ref{twistL}) for this
case. We have%
\[%
\begin{tabular}
[c]{|l|}\hline
$Z_{Aa}=\left(
\genfrac{}{}{0pt}{}{{8bose}}{{4fermi}}%
\right)  \;\;\;%
\genfrac{}{}{0pt}{}{\text{{12x2 rectangular matrix,\ }}{A=1,\cdots
,12;\;}\text{{\ a}}=1,2}{\text{{2 twistors in fundamental rep of }}%
{OSp(8}^{\ast}{|4)}}%
$\\\hline
$Z_{Aa}~=\text{{ (12,2)} { of }}{OSp(8}^{\ast}{|4)}_{\text{global}}%
\times\text{{ SU(2)}}_{\text{local}}$\\\hline
$L=\bar{Z}^{Aa}\left(  \left(  \partial+A\right)  Z\right)  _{Aa}%
\rightarrow\text{\ }\bar{Z}^{aA}Z_{Ab}=0,~~%
\genfrac{}{}{0pt}{}{\text{SU(2)~gauge invariants}}{\text{{in Fock space }%
}~~~~~~~~~~~}%
$\\\hline
$%
\genfrac{}{}{0pt}{}{\text{{Pseudo-real~~~~~~~~}}}{\text{1st \&~2nd~related}}%
\text{\ }Z_{Aa}=\left(
\begin{array}
[c]{cc}%
a_{{1}i} & a_{{2}i}\\
\bar{a}_{{2}}^{i} & -\bar{a}_{{1}}^{i}\\
\psi_{{1}\alpha} & \psi_{{2}\alpha}\\
\bar{\psi}_{{2}}^{\alpha} & -\bar{\psi}_{{1}}^{\alpha}%
\end{array}
\right)  \;%
\genfrac{}{}{0pt}{}{i:~~4~of~~\text{SU}(4)\subset\text{SO}(6,2)}{\alpha
:\;~~2~of~~\text{SU}(2)\subset\text{Sp}(4)}%
$\\\hline
\end{tabular}
\ \
\]
Here $\bar{Z}^{Aa}$ is obtained from $Z_{Aa}$ by taking hermitian conjugation
and multiplying by the 12$\times12$ diagonal matrix $diag\left(  1_{4}%
,-1_{4},1_{2},1_{2}\right)  $. However, $Z_{Aa}$ is pseudo real, $\bar{Z}%
^{aA}=C^{AB}Z_{Bb}\varepsilon^{ba},$ as it is defined as part of the group
element $g\in$OSp$\left(  8^{\ast}|4\right)  $ with the correct signature.
Then $Z_{Aa}$ takes the form above in a natural basis. Thus the second column
is related to the first one, but still consistent with a local SU$\left(
2\right)  $ applied on $a=1,2.$ When $Z,\bar{Z}$ of these forms are inserted
in the Lagrangian, it is seen that according to the canonical formalism, the
oscillators identified above all have positive norm $\left[  a_{1i},a_{1}%
^{j}\right]  =\delta_{i}^{j}=\left[  a_{2i},a_{2}^{j}\right]  $, $\left\{
\psi_{{1}\alpha},\psi_{{1}}^{\beta}\right\}  =\delta_{\alpha}^{\beta}=\left\{
\psi_{{2}\alpha},\psi_{{2}}^{\beta}\right\}  .$ We count the degrees of
freedom before the constraints, and find that $Z_{Aa}$ has (8$_{B}$+4$_{F}%
$)x2$_{\text{(complex)}}$=16$_{B}+8_{F}$ (namely $a_{{1}i},a_{{2}i},\psi
_{{1}\alpha},\psi_{{2}\alpha}$). The constraints are due to a SU$\left(
2\right)  $ gauge symmetry acting on the index $a=1,2$ (although it seems like
SU$\left(  2\right)  \times$U$\left(  1\right)  ,$ the U$\left(  1\right)  $
part is automatically satisfied because of the pseudoreal form of $Z_{Aa}$).
The 3 gauge parameters and 3 constraints remove 6 bosonic degrees of freedom,
and we remain with 10$_{B}+8_{F}$ physical degrees of freedom. This is the
same as the count for the superparticle ($5x,5p,8\theta$). It is obvious we
have the same number of degrees of freedom and the same symmetries OSp$\left(
8^{\ast}|4\right)  ,$ with the symmetry being much more transparent in the
twistor basis.

Instead of constrained twistors we can also express this theory in terms of
unconstrained coset parameters in the form (\ref{ht1},\ref{ht2}) where
$h_{\Gamma}$ was given in \cite{2tsuperstring}, with more details in
\cite{twistorD}.

The quantum theory can proceed in terms of twistors or in terms of constrained
oscillators. The resulting representation, after satisfying the SU$\left(
2\right)  $ constraints in the Fock space of the oscillators defined above, is
precisely the doubleton of OSp$\left(  8^{\ast}|4\right)  ,$ and this is
precisely equivalent to the fields in Eq. (\ref{6Dfields}). This oscillator
representation was worked out a long time ago in \cite{gunaydinWarner} using
again the BG method \cite{barsgunaydin}. The selection of the doubleton among
many other Fock space states discussed in \cite{gunaydinWarner} is the analog
of choosing the SU$\left(  2\right)  $ \textquotedblleft color" singlet in
analogous discussion to the one in \cite{2tZero}. More details will be given
in \cite{twistorD}.

\subsection{Supertwistors for d=10 and AdS$_{5}\times$S$^{5}$ KK towers}

This was explained in detail in \cite{2tAdSs}. Here we will quickly count
degrees of freedom for the AdS$_{5}\times$S$^{5}$ superparticle. This
superparticle starts out with $10x,10p,16\theta$ real degrees of freedom.
Fixing $\tau$ gauges and solving constraints, reduces the physical degrees of
freedom down to 9$x$, 9$p$, 16 $\theta.$ With 16$\theta$'s we construct the
Clifford algebra that is realized on states with 128$_{B}+128_{F}.$ These
correspond to the supergravity multiplet. Hence this case is related to gravity.

The superparticle action has a hidden global superconformal symmetry
PSU$\left(  2,2|4\right)  $ whose generators are given in \cite{2tAdSs},
therefore the physical states should be classified as a unitary representation
under this group. The resulting spectrum coincides with all the infinite
Kaluza-Klein towers of type IIB supergravity compactified on AdS$_{5}\times
$S$^{5}.$

Now we turn to the twistor gauge. This is a different gauge choice compared to
(\ref{twistgauge1}). We have split phase space into two groups of six each, as
$X^{M}=\left(  X^{m},X^{I}\right)  ,$ $P^{M}=\left(  P^{m},P^{I}\right)  ,$
but the Sp$\left(  2\right)  $ constraints is SO$(10,2)$ covariant for the
overall 12 dimensions $X^{2}=P^{2}=X\cdot P.$ Using the Sp$\left(  2\right)
\times$SO$\left(  4,2\right)  \times$SO$\left(  6\right)  $ gauge symmetries
we choose gauges and solve all the Sp$\left(  2\right)  $ constraints with the
following configuration
\begin{align}
M  &  =\left(  \,\,0^{\prime}~~0~~~1~\cdots~4~~,~~\;\;\;5\,~~\;6~~~\;7~~\cdots
~10\right) \nonumber\\
X^{M}  &  \sim\left(  \,\,a~~~0~~~0~~\cdots~0~~,~~~~a\,~~~0~~~~0~~\cdots
~0\right) \label{asdgauge}\\
P^{M}  &  \sim\left(  \,\,0~~~b~~~0~~\cdots~0~~,~~~~0\,~~~b~~~~0~~\cdots
~0\right) \nonumber
\end{align}
In this configuration the only nonzero angular momenta that couple according
to the scheme given above $L^{0^{\prime}0}=ab\equiv l$ and $L^{56}=ab\equiv
l.$ Therefore, instead of Eqs. (\ref{twistJ},\ref{twistL}) we now obtain, with
$g\in$PSU$\left(  2,2|4\right)  ,$%
\begin{align}
L  &  =-\frac{l}{{2}}Str\left(  \partial_{\tau}gg^{-1}\hat{\Gamma}\right)
,\;\;J_{{A}}^{~{B}}\sim\left(  g^{-1}\hat{\Gamma}g\right)  _{{A}}^{~{B}%
}\;\;,\;\;\hat{\Gamma}\equiv\left(
\genfrac{}{}{0pt}{}{{\Gamma}^{{0}^{\prime}0}}{0}%
\genfrac{}{}{0pt}{}{0}{-\Gamma^{56}}%
\right)  ,\label{AdScoset}\\
g  &  \in\text{{ PSU}}{(2,2|4)}\text{{~/~[PSU}}{(2|2)}\times\text{{PSU}%
}{(2|2)}\text{{]}}\nonumber
\end{align}
In an appropriate basis ${\Gamma}^{{0}^{\prime}0},\Gamma^{56}$ can be taken to
be diagonal 4$\times4$ matrices, each with two $+1$ and two $-1$ eigenvalues,
therefore $\hat{\Gamma}$ is the diagonal matrix $diag\left(  1_{2}%
,-1_{2},-1_{2},1_{2}\right)  .$ Any generator that commutes with this matrix
is a remaining gauge symmetry. Thus, there is still the gauge symmetry {[PSU(2%
$\vert$%
2)xPSU(2%
$\vert$%
2)]. The first PSU(2%
$\vert$%
2) acts on rows }$1,2,7,8$ of $g$ ($+1$ eigenvalues of $\hat{\Gamma}$) while
the second {PSU(2%
$\vert$%
2) acts on rows }$3,4,5,6$ ($-1$ eigenvalues of $\hat{\Gamma}$). After
removing the gauge degrees of freedom the $g$ in Eq.(\ref{AdScoset}) belongs
only to the coset $g\in${ PSU(2,2%
$\vert$%
4)~/~[PSU(2%
$\vert$%
2)xPSU(2%
$\vert$%
2)]}. We count the number of physical degrees of freedom as follows. The full
PSU$\left(  2,2|4\right)  $ supergroup contains $15+15$ real bosons and 16
complex fermions, thus altogether $30_{B}+32_{F}.$ The gauge subgroup {PSU(2%
$\vert$%
2) contains 3+3 bosons and 4 complex fermions. Therefore PSU(2%
$\vert$%
2)xPSU(2%
$\vert$%
2) has 12}$_{B}+16_{F}$ real gauge degrees of freedom. The physical degrees of
freedom in the coset is the difference, namely $18_{B}+16_{F}.$ As expected
this is the same number as the 9$x$, 9$p$, 16 $\theta$ we counted for the
AdS$_{5}\times$S$^{5}$ superparticle above. The hidden SU$\left(
2,2|4\right)  $ symmetries of the superparticle are evident in this twistor
gauge. Hence, one alternative description of AdS$_{5}\times$S$^{5}$ is the
coset given above. This is a new observation.

We can now rewrite this in terms of constrained twistors, as we did for
$d=4,6$ above. Rather than removing all of the gauge degrees of freedom we
allow some of it to remain. Then from $g$ we can extract four twistors which
we call $Z_{Aa}$ with the following properties\footnote{These correspond to
the four middle columns of $g^{-1},$ or equivalently the first two and last
two columns of $g^{-1}$, as discussed in \cite{2tsuperstring}.}%
\begin{equation}%
\begin{tabular}
[c]{|l|}\hline
$\overset{{A=1,\cdots,8}}{Z_{Aa}=}(%
\genfrac{}{}{0pt}{}{\overset{a=1,2}{{bose}}}{{fermi}}%
\genfrac{}{}{0pt}{}{\overset{a=3,4}{{fermi}}}{{bose}}%
)\;\;\;%
\genfrac{}{}{0pt}{}{\text{{8x4 rectangular matrix\ \ \ \ \ \ \ \ \ \ \ }%
}}{\text{{4 fundamental reps of PSU(2,2$|$4)}}}%
\genfrac{}{}{0pt}{}{\text{{ }}}{^{\text{{ }}}}%
$\\\hline
$Z_{Aa}~=\text{{ (8,4)} { of }}{PSU(2,2|4)}_{\text{\emph{global}}}%
\times{[PSU(2|2)\times U(1)]}_{\text{\emph{local}}}$\\\hline
$L=\bar{Z}^{Aa}\left(  \left(  \partial+A\right)  Z\right)  _{Aa}\rightarrow
$\ $\bar{Z}^{aA}Z_{Ab}=0,~~_{\text{{ }}}%
\genfrac{}{}{0pt}{}{\text{take}~\text{{gauge invariants}}}{\text{{in Fock
space~~~~~~~~~~}}}%
$\\\hline
\end{tabular}
\ \ \label{AdsTwistors}%
\end{equation}
The first two twistors $a=1,2$ each has four bosons in $4$ of SU$\left(
2,2\right)  $ and four fermions in $4$ of SU$\left(  4\right)  .$ The last two
twistors $a=3,4$ each has four fermions in $4$ of SU$\left(  2,2\right)  $ and
four bosons in $4$ of SU$\left(  4\right)  .$ On the basis $a=1,2,3,4$ we act
with a gauge symmetry ${PSU(2|2)\times U(1),}$ hence the Lagrangian is
invariant under ${PSU(2,2|4)}_{\text{\emph{global}}}\times{[PSU(2|2)\times
U(1)]}_{\text{\emph{local}}}.$ Let us count the degrees of freedom. The
complex $Z_{Aa}$ has (16$_{B}$+16$_{F}$)x2$_{\text{(complex)}}$, therefore
$32_{B}+32_{F}$ real degrees of freedom. PSU(2%
$\vert$%
2)$\times$U(1) has (3+3+1)$_{B}$+8$_{F}$ real gauge parameters. The gauge
parameters together with the corresponding constraints remove $14_{B}+16_{F}.$
Therefore the physical degrees of freedom in $Z_{Aa}~$is the difference,
namely $18_{B}+16_{F},$ which is the same as the 9$x$, 9$p$, 16 $\theta$ we
counted for the AdS$_{5}\times$S$^{5}$ superparticle above. The global
symmetry is still ${PSU(2,2|4),}$ and it has become evident rather than hidden
in the twistor version we have just described. Hence, a new alternative
description of AdS$_{5}\times$S$^{5}$ is the constrained twistors $Z_{Aa}$
given above.

The quantum theory for this case can again be described by using the BG
oscillator approach \cite{barsgunaydin}\cite{2tZero}. But now the
\textquotedblleft color" group is the supergroup ${PSU(2|2)\times U(1),}$ and
the discussion in \cite{2tZero} should be modified accordingly. The
\textquotedblleft color" supergroup is mathematically a new case in the BG
approach, and will be further analyzed in \cite{twistorD}.

\subsection{Bosonic twistors in any $d$}

The methods above can be applied in any dimension with the purely bosonic
group $G=$SO$\left(  d,2\right)  $ as listed in Eq.(\ref{table1}). In this
case the analogs of the twistors $Z_{Aa}$ correspond again to 1/4 of the
columns of $g^{-1}$. But for sufficiently large $d$ (namely $d>6$) there are
more entries $Z_{Aa}$ than there are group parameters in $SO\left(
d,2\right)  ;$ hence these $Z_{Aa}$ come out with lots of constraints among
the entries in these rectangular matrices, as dictated by the spinor
representation of the group SO$\left(  d,2\right)  $. The independent
parameters correspond to the coset $t_{\Gamma}$ as in Eqs. (\ref{ht1}%
,\ref{ht2}). These give the correct generalizations of twistors in the sense
that they provide an alternative (twistor) description of the massless
relativistic particle in $d$ dimensions. It is a holographic image in the 2T
structure. As $d$ gets larger and larger beyond $d=6$ it becomes cumbersome to
try to describe these in terms of oscillators, because of the complexity of
the constraints on the oscillators. However, if we relax the requirement of
only particles, and admit also D-brane degrees of freedom, as in footnote
(\ref{Gbose}), then the oscillator (or twistor) approach becomes again a very
efficient tool. The oscillator version including D-branes was described in
\cite{2ttwistor} for the group OSp$\left(  M|2N\right)  ,$ and applied to the
toy M-model for OSp$\left(  1|64\right)  $ in \cite{2ttoyM}. A more detailed
discussion of the twistors for general $d,$ without and with $D$-branes, will
appear in \cite{twistorD}.

\section{2T superstrings d=3,4,5,6,10}

So far in this lecture we discussed superparticles and the associated
supertwistors, and their physical spectra. These have a direct generalization
to superstrings via the 2T superstring formalism given in \cite{2tsuperstring}%
. Briefly, the action is

\[
L_{2T}^{\pm}=%
\genfrac{}{}{0pt}{}{{\partial}_{\pm}{X\cdot P}^{\pm}{-}\frac{1}{2}{AX\cdot
X-}\frac{1}{2}{B}_{\pm\pm}{P}^{\pm}{\cdot P}^{\pm}{-C}_{\pm}{P}^{\pm}{\cdot
X}}{-\frac{1}{2^{[d/2]-1}}Str\left(  \partial_{\pm}g\bar{g}\left(
\genfrac{}{}{0pt}{}{L_{MN}^{\pm}\Gamma^{MN}}{0}\genfrac{}{}{0pt}{}{0}{0}%
\right)  \right)  +\mathit{L}_{G}}%
\]
$X_{M}(\tau,\sigma),P_{M}^{\pm}(\tau,\sigma),$\ $L_{MN}^{\pm}=X_{[M}%
P_{M]}^{\pm},~g(\tau,\sigma)$ are now string fields, and ${\partial}_{\pm
}=\frac{1}{2}\left(  \partial_{\sigma}\pm\partial_{\tau}\right)  .$ Here
$L_{2T}^{\pm}$ represent left/right movers, and similar to \cite{berko} there
is open string boundary conditions$.$ $L_{G}$ is an additional degree of
freedom that describes an internal current algebra for some SYM group $G.$ The
local and global symmetries are similar to those of the particle and are
described in \cite{2tsuperstring}. The global symmetry is $G_{d}$ chosen for
various $d$ as before, G$_{{d}}$=OSp(8%
$\vert$%
4)$_{{3}}$,SU(2,2%
$\vert$%
4)$_{{4}}$,F(4)$_{{5}}$,OSp(8%
$\vert$%
4)$_{{6}}.$ The particle gauge for these give usual type superstrings and the
twistor gauge gives twistor superstrings, with the twistors described above.
In the 2T philosophy each one of these have many duals that can be found and
investigated by choosing various gauges. This is a completely open field of
investigation at this time.

Similarly to the $d=10$ particle case we also consider the d=10 2T superstring

\[
L_{2T}^{\pm}=%
\genfrac{}{}{0pt}{}{{\partial}_{\pm}{\hat{X}\cdot\hat{P}}^{\pm}{-}\frac{1}%
{2}{A\hat{X}\cdot\hat{X}-}\frac{1}{2}{B}_{\pm\pm}{\hat{P}}^{\pm}{\cdot\hat{P}%
}^{\pm}{-C}_{\pm}{\hat{P}}^{\pm}{\cdot\hat{X}}}{-\frac{1}{8}Str\left(
\partial_{\pm}g\bar{g}\left(  \genfrac{}{}{0pt}{}{L_{MN}^{\pm}\Gamma^{MN}%
}{0}\genfrac{}{}{0pt}{}{0}{-L_{IJ}^{\pm}\Gamma^{IJ}}\right)  \right)  }%
\]
where SO(10,2)$\rightarrow$SO(4,2)$\times$SO(6), $\hat{X}^{M}=\left(  {X}%
^{m},{X}^{I}\right)  $, $\hat{P}_{M}^{\pm}=\left(  {P}_{m}^{\pm},{P}_{I}^{\pm
}\right)  $, $\ g(\tau,\sigma)\in SU\left(  {2,2|4}\right)  $, and
$L_{MN}^{\pm}=X_{[M}P_{M]}^{\pm}$, $L_{IJ}^{\pm}=X_{[I}P_{J]}^{\pm}$. The
local and global symmetries are discussed in detail in \cite{2tAdSs}%
,\cite{2tsuperstring}. In the particle-type gauge, the spectrum in the
particle limit is the same as linearized type IIB SUGRA compactified on
AdS$_{5}\times$S$^{5}.$ In the twistor gauge this theory is currently being
investigated by using the twistors in Eq. (\ref{AdsTwistors}). As usual ,
being a 2T theory we expect dual versions of the theory in other gauges. This
could be very interesting in terms of M-theory.

\section{Closing Remarks}

I have described the following established facts about 2T-physics

\begin{itemize}
\item 2T-physics, with local Sp(2,R) \& generalizations, gives emergent
dynamics/spacetimes via d+2 $\rightarrow$d holography.

\item 1T-physics corresponds to d-dimensional holographs of the $d+2$ theory.
Various holographs are dual; and the\ duality group is Sp$(2,R)$ \& generalizations.

\item When $d+2$ is in flat space each holograph has hidden SO$(d,2)$
symmetry. Its quantum Hilbert space forms a basis for same eigenvalues of the
Casimirs of SO$(d,2)$. This applies with or without spin or supersymmetry.

\item Twistor space is a particular hologram of the $d+2$ theory. In the
twistor gauge the SO$\left(  d,2\right)  $ becomes more manifest as compared
to other holograms.

\item The $10+2$ twistor string, and its particle version show hidden $(10,2)$
holographic structures in AdS$_{5}\times$S$^{5}$ superstring and supergravity.
These are strong indications that other aspects of $M$ theory also have a 2T
description. See related remarks in \cite{2tsuperstring} that connect to
\cite{vasiliev}\cite{bianchi}.
\end{itemize}

At this point it is hard to resist to also make some speculations, as follows

\begin{itemize}
\item The currently known corners of M-theory are very likely holograms of the
same nature. The known M-dualities appear to be analogs of the Sp(2,R) \&
generalizations. This provides hints for the underlying gauge symmetry.

\item Together with the earlier indications described in the introduction, it
seems now even more likely that M-theory would be most clearly formulated as a
13D theory with signature $(11,2)$ and global supersymmetry OSp$(1|64)$.
\end{itemize}

\section{Appendix: twistor string with SO$\left(  3,1\right)  $ signature}

Many people in this conference raised questions on the analytic continuation
of SO$\left(  3,1\right)  $ to SO$\left(  2,2\right)  $ in the twistor
superstring. Since I gave some thought to this point, I outline below what the
differences are when one uses the correct signature SO$\left(  3,1\right)  .$
There are definite \textit{changes in the formulation of the theory}, beyond
the naive analytic continuation that inserts $i$ in appropriate places, as
follows. I work in the Berkovits formulation as it appears in \cite{berkWit},
since this is the form of the 2T theory when we gauge fix the 2T superstring
to the twistor gauge \cite{2tsuperstring}.

When the signature is (3,1) the twistor $Y^{A}$\ in \cite{berkWit} must be the
complex conjugate of the twistor $Z_{A},$\ except for the metric, namely
$Y^{A}=\bar{Z}^{A}$ as seen in Eq. (\ref{4Dtwist}). Therefore, $Z_{A}$\ and
$\bar{Z}^{A}$\ must have the same worldsheet conformal dimension $1/2.$\ This
differs from Berkovits's $\dim\left(  Z\right)  =0$\ and $\dim\left(
Y\right)  =1.$\ There is a definite consequence: There must be a shift in the
stress tensor $T$, and in all the dimensions of the wavefunctions of physical
vertices$,$ as remarked in \cite{2tsuperstring}. The following table gives the shifts%

\[%
\begin{tabular}
[c]{|l|l|l|}\hline
signature & \ $\;\;\;\left(  2,2\right)  $ & $\;\ \;\;\;\;\;\;$\ $\;\left(
3,1\right)  $\\\hline
$\underset{\text{dimensions\ \ \ \ \ \ \ \ \ }}{\text{stress tensor}}$ &
$\underset{2\;\;\;~\;\;1\;\;\;1\;\;\;0}{T=Y^{A}\partial Z_{A}}$ &
$\underset{2\;\;}{\tilde{T}=}\frac{1}{2}\underset{~1/2\;\;1\;\;1/2}{\bar
{Z}^{A}\partial Z_{A}}-\frac{1}{2}\partial\bar{Z}^{A}Z_{A}$\\\hline
$\underset{\text{dimensions\ \ \ \ \ \ \ \ \ }}{\text{SYM vertex op}}$ &
$\underset{1\;\;\;\;\;\;1\;\;\;0\;\;\;\;\;\;\;}{V_{\phi}=j^{a}\phi_{a}\left(
Z\right)  }$ & $\underset{1\;\;\;\;\;\;1\;\;\;0\;\;Z\leftrightarrow
tZ\;\dim~1/2\;\text{OK}}{V_{\phi}=j^{a}\phi_{a}\left(  Z\right)  \;\;\text{no
changes}}$\\\hline
$\underset{\text{helicity +2, dims\ \ \ \ \ \ \ \ \ }}{\text{Conf. SUGRA}}$ &
$\underset{1\;\;\;\;\;\;\;1\;\;\;0\;\;\;\;\;}{V_{f}=Y^{A}f_{A}\left(
Z\right)  }$ & $\underset{1\;\;\;\;\;\;1/2\;\;1/2\;\;Z\leftrightarrow
tZ\;}{V_{f}=\bar{Z}^{A}f_{A}\left(  Z\right)  \;\;\;}$\\\hline
$\underset{\text{helicity -2, dims\ \ \ \ \ \ \ \ \ }}{\text{Conf. SUGRA}}$ &
$\underset{1\;\;\;\;\;\;1\;\;0\;\;0\;\;\;\;\;}{V_{g}=\partial Z_{A}%
g^{A}\left(  Z\right)  }$ & $\underset{1\;\;\;\;\;\;\;1\;\;\frac{1}{2}%
\;-\frac{1}{2}\;\;Z\leftrightarrow tZ\;\;\;\;\;}{V_{g}=\partial Z_{A}%
g^{A}\left(  Z\right)  \;\;\;\;\;\;}$\\\hline
$\underset{\text{instanton \ number\ }}{\text{Amplitudes\ }}$ & $\underset
{n}{T_{n}=T+n\partial\left(  YZ\right)  }$ & $\underset{=T_{n}-\frac{1}%
{2}\partial\left(  YZ\right)  \;\leftrightarrow\text{\ }n\rightarrow
n-1/2}{\tilde{T}_{n}=\tilde{T}+n\partial\left(  \bar{Z}Z\right)
\;\;\;\;\;\;\;\;}$\\\hline
\end{tabular}
\ \ \
\]
Note that the SYM vertex $\phi_{a}\left(  Z\right)  $ has still dimension
zero, but now it must be constructed from $Z$ that has dimension $1/2$ instead
of $Z$ that has dimension $0.$ However since $\phi_{a}\left(  Z\right)  $ is
homogeneous it means it is constructed from ratios of components of $Z_{A}.$
Therefore the same wavefunctions will still appear, and so it seems that
nothing changes in the SYM sector. This is good news.

Although the SYM wavefunction $\phi_{a}\left(  Z\right)  $ has the same
dimension for either SO$\left(  3,1\right)  $ or SO$\left(  2,2\right)  $, the
conformal supergravity wavefunctions $f_{A}\left(  Z\right)  ,g^{A}\left(
Z\right)  $ must have different dimensions, as shown in the table. I had hoped
that $f_{A}(Z)$\ with $\dim(f)=\frac{1}{2}$\ and $g^{A}(Z)$\ with
$\dim(g)=-\frac{1}{2}$ would not exist (and therefore get rid of the
\textquotedblleft conformal gravity pollution" in the theory); but apparently
they do exist without much modification also.

Another change is the stress tensor itself. The shift is equivalent to a shift
in the instanton number $n\rightarrow n-1/2$. I have not checked the details
if this is cancelled by additional modifications, and whether this is harmless
as well.


\begin{thebibliography}{99}                                                                                               %


\bibitem {2treviews}I. Bars, C. Deliduman and O. Andreev, \textquotedblleft%
\ Gauged Duality, Conformal Symmetry and Spacetime with Two Times" , Phys.
Rev. \textbf{D58} (1998) 066004 [arXiv:hep-th/9803188]. For reviews of
subsequent work see: I. Bars, \textquotedblleft\ Two-Time Physics" , in the
Proc. of the 22nd Intl. Colloq. on Group Theoretical Methods in Physics, Eds.
S. Corney at. al., World Scientific 1999, [arXiv:hep-th/9809034];
\textquotedblleft\ Survey of two-time physics,"
Class.\ Quant.\ Grav.\ \textbf{18}, 3113 (2001) [arXiv:hep-th/0008164];
\textquotedblleft\ 2T-physics 2001,"\ AIP Conf. Proc. \textbf{589} (2001),
pp.18-30; AIP Conf. Proc. \textbf{607} (2001), pp.17-29 [arXiv:hep-th/0106021].

\bibitem {2ttwistor}I. Bars, \textquotedblleft\ 2T physics formulation of
superconformal dynamics relating to twistors and supertwistors,"
Phys.\ Lett.\ B \textbf{483}, 248 (2000) [arXiv:hep-th/0004090].

\bibitem {twistorD}I. Bars and M. Picon, in preparation.

\bibitem {witten}E. Witten, \textquotedblleft Perturbative gauge theory as a
string theory in twistor space", Commun. Math. Phys. \textbf{252} (2004) 189
[arXiv:hep-th/0312171]; \textquotedblleft Parity invariance for strings in
twistor space", hep-th/0403199.

\bibitem {witten2}F. Cachazo, P. Svrcek and E. Witten, \textquotedblleft\ MHV
vertices and tree amplitudes in gauge theory", JHEP \textbf{0409} $\left(
2004\right)  $ 006 [arXiv:hep-th/0403047]; \textquotedblleft\ Twistor space
structure of one-loop amplitudes in gauge theory", JHEP \textbf{0410} $\left(
2004\right)  $ 074 [arXiv:hep-th/0406177]; \textquotedblleft Gauge theory
amplitudes in twistor space and holomorphic anomaly", JHEP \textbf{0410}
$\left(  2004\right)  $ 077 [arXiv:hep-th/0409245]

\bibitem {berko}N. Berkovits, ``An Alternative string theory in twistor space
for N=4 Super Yang-Mills", Phys. Rev. Lett. \textbf{93} (2004) 011601 [arXiv:hep-th/0402045].

\bibitem {berko2}N. Berkovits and L. Motl, \textquotedblleft Cubic twistorial
string field theory", JHEP \textbf{0404} (2004) 56 [arXiv:hep-th/0403187].

\bibitem {berkWit}N. Berkovits and E. Witten, \textquotedblleft Conformal
supergravity in twistor-string theory", JHEP \textbf{0408} (2204) 009 [arXiv:hep-th/0406051].

\bibitem {2tsuperstring}I. Bars, \textquotedblleft Twistor superstring in
2T-physics,\textquotedblright\ Phys. Rev. \textbf{D70} (2004) 104022 [arXiv:hep-th/0407239].

\bibitem {2tAdSs}I. Bars, \textquotedblleft\ Hidden 12-dimensional structures
in AdS$_{5}$ x S$^{5}$ and M$^{4}$ x R$^{6}$ supergravities," {}Phys.\ Rev.\ D
\textbf{66}, 105024 (2002) [arXiv:hep-th/0208012].

\bibitem {2tZero}I. Bars, \textquotedblleft\ A mysterious zero in AdS$_{(}5)$
x S$^{5}$ supergravity," Phys.\ Rev.\ D \textbf{66}, 105023 (2002) [arXiv:hep-th/0205194].

\bibitem {barsgunaydin}I. Bars and M. G\"{u}naydin, Comm. Math. Phys.
\textbf{91 }(1983) 31.

\bibitem {gunaydinWarner}M. G\"{u}naydin, P. van Nieuwenhuizen, N.P. Warner,
Nucl. Phys. \textbf{B255} (1985) 63.

\bibitem {witten6D}E. Witten, \textquotedblleft Conformal field theory in four
and six dimensions", In \textit{Oxford 2002, Topology, geometry and quantum
field theory}, p.405.

\bibitem {osaka}I. Bars, \textquotedblleft Duality and Hidden Dimensions",
hep-th/9604200, published in Frontiers in Quantum Field Theory, Eds. H.
Itoyama et. al., World Scientific (1996), page 52. See also, \textquotedblleft
Supersymmetry, P-Brane Duality And Hidden Space-Time Dimensions", Phys. Rev.
\textbf{D54} (1996) 5203 [=hep-th/9604139].

\bibitem {stheory}I. Bars, S-theory, Phys. Rev. \textbf{D55} (1997) 2373
[arXiv:hep-th/9607112]; \textquotedblleft Algebraic structure of S-theory",
hep-th/9608061; \textquotedblleft Black hole entropy reveals a tweflth
dimension", Phys. Rev. D55 (1997) 3633 [arXiv:hep-th/9610074];
\textquotedblleft A Case for fourtheen dimensions", Phys. Lett. \textbf{B403}
(1997) 257 [arXiv:hep-th/9704054]

\bibitem {ftheory}C. Vafa, \textquotedblleft Evidence for F-Theory", Nucl.
Phys. \textbf{B469} (1996) 403 [arXiv:hep-th/9602022]

\bibitem {ibkounnas}I. Bars and C. Kounnas, \textquotedblleft Theories with
two times", Phys. Lett. \textbf{B402} (1997) 25 [arXiv:hep-th/9703060];
\textquotedblleft String and particle with two times", Phys. Rev. \textbf{D56}
(1997) 3664 [arXiv:hep-th/9705205]; I. Bars and C. Deliduman,
\textquotedblleft Superstrings with new supersymmetry in (9,2)-dimensions and
(10,2)-dimensions", Phys. Rev. \textbf{D56} (1997) 6579,
[arXiv:hep-th/9707215]; \textquotedblleft Gauge principles for multi -
superparticles", Phys. Lett. \textbf{B417} (1998) 240 [arXiv:hep-th/9710066]

\bibitem {liftM}I. Bars, C. Deliduman and D. Minic, \textquotedblleft\ Lifting
M-theory to Two-Time Physics", Phys. Lett. \textbf{B457} (1999) 275 [arXiv:hep-th/9904063].

\bibitem {2ttoyM}I. Bars, \textquotedblleft Toy M-model", unpublished. For a
brief outline of the model see \cite{liftM}\cite{2ttwistor} and third
reference in \cite{2treviews}.

\bibitem {2tHandAdS}I. Bars, \textquotedblleft Conformal symmetry and duality
between free particle, H-atom and harmonic oscillator", Phys. Rev.
\textbf{D58} (1998) 066006 [arXiv:hep-th/9804028]; \textquotedblleft Hidden
Symmetries, AdS$_{d}\times S^{n}$, and the lifting of one-time physics to
two-time physics", Phys. Rev. \textbf{D59} (1999) 045019 [arXiv:hep-th/9810025].

\bibitem {2tbacgrounds}I. Bars, \textquotedblleft Two time physics with
gravitational and gauge field backgrounds", Phys. Rev. \textbf{D62}, 085015
(2000) [arXiv:hep-th/0002140]; I. Bars and C. Deliduman, \textquotedblleft%
\ High spin gauge fields and two time physics", Phys. Rev. \textbf{D64},
045004 (2001) [arXiv:hep-th/0103042].

\bibitem {2tspinning}I. Bars and C. Deliduman, \textquotedblleft Gauge
symmetry in phase space with spin: a basis for conformal symmetry and duality
among many interactions", Phys. Rev. \textbf{D58} (1998)106004,
[arXiv:hep-th/9806085]. For additional quantum properties of the spinning
theory in covariant quantization, including interactions, see the first
reference in \cite{2tfield}.

\bibitem {2tSuper}I. Bars, C. Deliduman and D. Minic, \textquotedblleft%
\ Supersymmetric two-time physics" , Phys. Rev. \textbf{D59 }(1999) 125004 [arXiv:hep-th/9812161].

\bibitem {2tString}I. Bars, C. Deliduman and D. Minic, \textquotedblleft%
\ Strings, Branes and Two-Time Physics" , Phys. Lett. \textbf{B466} (1999) 135 [arXiv:hep-th/9906223].

\bibitem {2tfield}I. Bars, \textquotedblleft\ Two-time physics in Field
Theory" , Phys.\ Rev.\ D \textbf{62}, 046007 (2000), [arXiv:hep-th/0003100];
\textquotedblleft U*(1,1) noncommutative gauge theory as the foundation of
2T-physics in field theory," Phys.\ Rev.\ D \textbf{64}, 126001 (2001)
[arXiv:hep-th/0106013]; I.~Bars and S.~J.~Rey, \textquotedblleft%
\ Noncommutative Sp(2,R) gauge theories: A field theory approach to two-time
physics," Phys.\ Rev.\ D \textbf{64}, 046005 (2001) [arXiv:hep-th/0104135].

\bibitem {vasiliev}M.A. Vasiliev, \textquotedblleft\ Higher spin superalgebras
in any dimension and their representations" , arXiv:hep-th/0404124.

\bibitem {11Dstring}I. Bars, `` Stringy Evidence for D=11 Structure in
Strongly Coupled Type IIA Superstring" , Phys.Rev. \textbf{D52} (1995) 3567 [arXiv:hep-th/9503228].

\bibitem {bianchi}M.~Bianchi, J.~F.~Morales and H.~Samtleben, ``On stringy
AdS(5) x S**5 and higher spin holography,'' JHEP \textbf{0307} (2003) 062
[arXiv:hep-th/0305052]; N.~Beisert, M.~Bianchi, J.~F.~Morales and
H.~Samtleben, ``On the spectrum of AdS/CFT beyond supergravity,'' JHEP
\textbf{0402} (2004) 001 [arXiv:hep-th/0310292]; ``Higher spin symmetry and N
= 4 SYM,'' [arXiv:hep-th/0405057].
\end{thebibliography}
\end{document}